\documentclass[aps,prl,twocolumn,floatfix,superscriptaddress,amsmath,amssymb]{revtex4-1}
\usepackage{graphicx}
\usepackage{bm}
\usepackage[pdftex]{color}
\usepackage{xcolor}
\usepackage{hyperref}
\usepackage[mathlines]{lineno}
\usepackage{braket}
\usepackage{bbm}
\usepackage{verbatim}
\usepackage{ulem}
\usepackage{standalone}
\usepackage{comment}
\usepackage{dsfont}
\usepackage{slashed}
\usepackage{braket}
\usepackage{tikz}
\usepackage{tikz-cd}
\usepackage{courier}
\usepackage{bm}
\usepackage{xcolor}
\usepackage{mdframed}
\usepackage{datetime}
\usepackage{bbm}
\usepackage{multirow}
\usepackage{cancel}
\usepackage{caption}
\usepackage{subcaption}
\usepackage{graphicx}
\usepackage[compat=1.1.0]{tikz-feynman}
\usepackage{array} 
\usepackage{booktabs}
\usepackage{standalone}

\newcommand{\parR}[1]{\textbf{\textit{#1}}---}

\newcommand{\Z}{\mathbb{Z}}

\newtheorem{theorem}{Theorem}

\renewcommand\[{\begin{equation}}
\renewcommand\]{\end{equation}}

\newcommand{\Ucl}{\mathcal U_{\mathrm{cl}}}
\def\ie{\begin{equation}\begin{aligned}}
\def\fe{\end{aligned}\end{equation}}
\newcommand{\prlsection}[1]{\noindent\textbf{{#1}--- }}
\newcommand{\Deff}{D_{\mathrm{eff}}}
\newcommand{\ii}{\mathrm i}
\newcommand{\cJ}{\mathcal J}

\begin{document}

\title{Noninvertible Lieb–Schultz–Mattis Constraints and Deconfined Criticality}

%
%

\author{Linhao Li}
\email{linhaoli601@gmail.com}
\affiliation{Department of Physics$,$ The Pennsylvania State University$,$ University Park$,$ Pennsylvania$,$ 16802$,$ USA}

\author{Zhen Bi}
\email{zjb5184@psu.edu}
\affiliation{Department of Physics$,$ The Pennsylvania State University$,$ University Park$,$ Pennsylvania$,$ 16802$,$ USA}
\affiliation{Center for Theory of Emergent Quantum Matter$,$ Institute for Computational and Data Sciences$,$ The Pennsylvania State University$,$ University Park$,$ Pennsylvania$,$ 16802$,$ USA}

\date{\today}

\begin{abstract}
We derive Lieb–Schultz–Mattis theorems for one-dimensional qudit systems with a class of internal noninvertible symmetries implemented by bilinear phase map operators. We show that any Hamiltonian preserving these symmetries together with translation or inversion symmetry must either be gapless or have degenerate ground states. Unlike conventional LSM constraints based on projective representations, these new constraints arise from nontrivial algebraic relations between the internal and crystalline symmetries. We illustrate this mechanism in a $\mathbb{Z}_4$ qudit model hosting two competing phases that spontaneously break the internal noninvertible symmetry and inversion symmetry, respectively. Numerical simulations reveal a direct continuous transition between these phases, realizing a stable deconfined quantum critical point involving an internal noninvertible symmetry.

\end{abstract}
\maketitle

\parR{Introduction } A central question in condensed matter physics is how symmetries constrain the possible low-energy behavior of a quantum many-body system. The Lieb--Schultz--Mattis (LSM) theorem~\cite{Lieb:1961aa} and its generalizations~\cite{Affleck:1986aa,OYA1997,Oshikawa:2000aa,Hastings:2004ab,Chen-Gu-Wen_classification2010,Fuji-SymmetryProtection-PRB2016,Watanabe:2015aa,Ogata:2018aa,Ogata:2020aa,Yao:2021aa,Yao:2019aa,PhysRevB.110.045118,PhysRevB.106.224420,PhysRevLett.133.136705} provide a powerful answer. They show that the interplay between internal and crystalline symmetries can forbid a unique gapped ground state, enforcing either gapless excitations or ground-state degeneracy associated with spontaneous symmetry breaking or topological order. In familiar examples, the LSM constraint originates from a nontrivial projective representation of the internal symmetry carried by each unit cell. Phases that spontaneously break different symmetries can satisfy the same LSM constraint, providing a natural setting for deconfined quantum critical points (DQCPs)~\cite{PhysRevB.70.144407,senthil2004deconfined,PhysRevLett.98.227202,PhysRevX.7.031051,PhysRevB.98.085140,PhysRevB.100.125137,PhysRevB.99.075103,PhysRevB.99.165143,PhysRevLett.115.267203,PhysRevLett.130.026801}, which mediate direct continuous transitions between these phases beyond the conventional Landau paradigm. A canonical example is the transition between N\'eel antiferromagnetic order and valence-bond-solid order, which break spin-rotation and lattice symmetries, respectively.

Recent developments have extended symmetry beyond groups to noninvertible symmetries described by fusion categories~\cite{PhysRevLett.93.070601,Frohlich:2006ch,Gaiotto:2014kfa,Aasen:2016dop,Bhardwaj:2017xup,Tachikawa:2017gyf,Chang:2018iay,thorngren2019fusion,Aasen:2020jwb,Thorngren:2021yso,Moradi:2022lqp,Bhardwaj:2022maz,Moradi:2023dan,Bhardwaj:2023fca,Lam:2023xng,PRXQuantum.4.020357,PhysRevB.108.214429,Cao:2023doz,Bhardwaj:2024wlr,Chen:2025uno,Choi:2024rjm,Jia:2024bng,Jia:2024zdp,Jia:2024wnu,Meng:2024nxx,Aksoy:2025rmg,Saito:2025qrp,Bhardwaj:2024qrf,Warman:2024lir,doi:10.1142/S0217751X25480021,Seifnashri:2024dsd,Decoppet:2024htz,PhysRevLett.126.195701,PhysRevB.104.125418,10.21468/SciPostPhys.11.4.082,PhysRevB.110.035155,pace2024gauging,cao2024generating,kim2025noninvertible,lu2026generalized,ning2026matrix,qmcw-dbrs}, revealing new quantum phases~\cite{PhysRevLett.133.116601,SciPostPhys.18.1.032,li2024non,Cao:2025qhg,ng8b-sdt4,chatterjee2024quantum,bhardwaj2024lattice,lz71-yf6f,PhysRevX.15.011058} and unconventional critical phenomena~\cite{SciPostPhys.18.5.153,Chatterjee:2024ych,mz32-k1zk,8tsg-q9j6,ebisu2026understanding,zhang2026e_,ravindran2026ginzburg,chen2026spontaneous}. A prominent example is Kramers--Wannier (KW) duality, which can itself impose LSM-type constraints~\cite{Seiberg:2023cdc,Seiberg:2024gek}. In its standard lattice realization, however, translation enters the fusion algebra, so the duality is not a purely internal symmetry. More generally, noninvertible symmetries whose algebra involves translation do not necessarily impose an LSM obstruction, as illustrated by projective noninvertible symmetry algebras admitting gapped entangled ground states~\cite{pace2025spacetime}. These developments raise several questions. Can the interplay between \textit{internal} noninvertible and crystalline symmetries give rise to LSM constraints? Do these require a nontrivial projective representation of the internal symmetry in each unit cell, or can they arise from a different algebraic mechanism? Can these constraints support DQCPs between phases that spontaneously break an internal noninvertible symmetry and a crystalline symmetry, respectively?

In this Letter, we establish a new family of LSM theorems for $1d$ qudit chains with internal noninvertible symmetries realized by bilinear phase map dualities, together with lattice translation or inversion symmetry. Unlike conventional LSM constraints based on projective representations carried by unit cells, these constraints arise from \textit{nontrivial algebraic relations} between internal and crystalline symmetries. Guided by these constraints, we construct a $\mathbb{Z}_4$ qudit model and numerically demonstrate a DQCP separating phases that spontaneously break the internal noninvertible symmetry and inversion symmetry, respectively. We establish the stability of this transition and uncover entanglement-spectrum signatures of gapless symmetry-protected-topological order at criticality.



\parR{Bilinear phase map duality }
We begin with a brief summary of the bilinear phase map (BPM) duality~\cite{LiOshikawaYan2026}, which generalizes the familiar KW duality. For a $\mathbb Z_N$ qudit chain of length $L$, the BPM duality operator is defined in the computational basis by
\begin{equation}
 D_A=\frac{1}{\sqrt{N^L}}
 \sum_{\{\hat{s}_j\},\{s_k\}}
 \omega^{\sum_{j,k}\hat{s}_j A_{jk}s_k}
 |\{\hat{s}_j\}\rangle\langle\{s_k\}|,
 \label{eq:BPM}
\end{equation}
where $\omega=e^{2\pi i/N}$, $s_j,\hat{s}_j\in\mathbb Z_N$ label the original and dual qudit variables, respectively, and $A$ is an $L\times L$ matrix over $\mathbb Z_N$. The matrix $A$ specifies the mapping between the original and dual configurations, thereby determining the duality map.

For any kernel vector $f\in\ker A$, satisfying $Af=0\pmod N$, the configurations $\{s_j\}$ and $\{s_j+f_j\}$ are mapped to the same state under $D_A$. A nonzero kernel vector therefore makes the map noninvertible, since $D_A$ cannot distinguish configurations related by the corresponding shift. Defining the corresponding shift operator by $U_f|\{s_j\}\rangle=|\{s_j+f_j\}\rangle$, we have
\begin{equation}
 D_A U_f=D_A,
 \qquad f\in\ker A.
\end{equation}
The operators $U_f$ generate the associated invertible kernel symmetries. Similarly, vectors $\hat f\in\ker A^{\mathsf T}$ generate the dual kernel symmetries $\widehat U_{\hat f}$, satisfying $\widehat U_{\hat f}D_A=D_A$.

The reverse duality is given by $D_A^\dagger=D_{-A^{\mathsf T}}$. Under periodic boundary conditions (PBC), the duality operators obey the fusion rules
\begin{equation}
 \begin{aligned}
 D_A^\dagger D_A
 &=\sum_{f\in\ker A}U_f,
 &
 D_A D_A^\dagger
 &=\sum_{\hat f\in\ker A^{\mathsf T}}\widehat U_{\hat f}.
 \end{aligned}
 \label{eq:BPMfusion}
\end{equation}
These fusion rules show that composing the duality with its reverse yields a sum over kernel symmetries rather than the identity whenever the kernel is nontrivial. The kernels $\ker A$ and $\ker A^{\mathsf T}$ characterize the invertible symmetries of the original and dual chains, respectively.

As a familiar example, the conventional KW duality is recovered for $N=2$ by choosing $A_{jk}=\delta_{j,k}-\delta_{j+1,k}$. Both $\ker A$ and $\ker A^{\mathsf T}$ are generated by the uniform vector $f=(1,1,\ldots,1)^{\mathsf T}$, corresponding to the global $\mathbb Z_2$ spin-flip symmetry. However, the reverse KW duality differs by a lattice translation $T$, i.e., $D_{\text{KW}}^\dagger=TD_{\text{KW}}$. As a result, its fusion rule reads $D_{\text{KW}}^2=T^{-1}(1+U_f)$, where the appearance of translation precludes interpreting the KW duality as an internal symmetry.



In this Letter, we consider a class of \textit{three-site BPM dualities} specified by
\begin{equation}
 A_{j-1,j}=\kappa,\qquad
 A_{j,j}=\mu,\qquad
 A_{j+1,j}=-\kappa,
 \label{gen:A}
\end{equation}
with all other matrix elements vanishing. We consider $\gcd(\kappa,N)=1$ and $2\mu=0\pmod N$ \footnote{The condition $\gcd(\kappa,N)=1$ ensures that each kernel solution is uniquely determined by two initial values; see the discussion around Eq.~\eqref{gen:fdefinition}.}, so that $\mu=0$ for odd $N$, while $\mu=0,N/2$ are allowed for even $N$. The corresponding BPM duality induces the following transformations:
\begin{equation}
 D_AX_j=\widehat B_jD_A,\quad
 D_AB_j=\widehat X_jD_A,\quad
 B_j=Z_{j-1}^{\kappa}Z_j^{\mu}Z_{j+1}^{-\kappa}.
 \label{gen:intertwine}
\end{equation}
Here the clock and shift operators acting on qudits are
\begin{equation}
 Z_i=\sum_{s=0}^{N-1}\omega^s\ket{s}\bra{s}_i,\quad
 X_i=\sum_{s=0}^{N-1}\ket{s+1}\bra{s}_i,\quad
 \omega=e^{2\pi i/N}.
\end{equation}
By construction, $A$ is antisymmetric, $A^{\mathsf T}=-A\pmod N$, which implies $D_A^\dagger=D_A$. Unlike the conventional KW duality, this three-site BPM duality and its reverse are therefore implemented by the same operator, whose fusion rule $D_A^2=\sum_{f\in\ker A}U_f$ contains no lattice translation. Thus, $D_A$ can realize an \textit{internal noninvertible symmetry}.

\parR{Noninvertible LSM theorem in qubit chains}
We now investigate LSM constraints arising from the interplay between the three-site BPM duality and translation or inversion symmetry.

As a motivating example, consider $N=2$ and $\kappa=\mu=1$, corresponding to the well-studied three-spin duality~\cite{LTurban_1982,PhysRevB.29.519,PhysRevB.26.6334,p32z-v884}. For $L\in3\mathbb Z$, the kernel is generated by $f^1=(1,1,0,1,1,0,\ldots)^{\mathsf T}$ and $f^2=(0,1,1,0,1,1,\ldots)^{\mathsf T}$. The corresponding kernel symmetry is $\mathbb Z_2\times\mathbb Z_2$ for $L\in3\mathbb Z$, generated by 
\begin{equation}
 U_1=\prod_jX_{3j-2}X_{3j-1},~
 U_2=\prod_jX_{3j-1}X_{3j}.    
\end{equation}
Together with the BPM duality, these operators generate the $\operatorname{Rep}(D_8)$ fusion category symmetry. 

There are three distinct $\operatorname{Rep}(D_8)$ symmetry-protected topological (SPT) phases with a unique gapped ground state~\cite{PhysRevLett.133.116601,Cao:2025qhg}. Thus, the internal noninvertible symmetry alone does not impose an LSM obstruction. The situation changes upon imposing lattice translation. Although $[D_A,T]=0$, translation acts on the kernel symmetries through an order-three automorphism,
\begin{equation}
 T:\quad U_1\longrightarrow U_2
 \longrightarrow U_1U_2
 \longrightarrow U_1.
 \label{gen:translationkernel}
\end{equation}
This automorphism induces a triality among the three $\operatorname{Rep}(D_8)$ SPT phases~\cite{thorngren2019fusion,antinucci2025categorical},
\begin{equation}
 T:\quad \mathrm{SPT}_1
 \longrightarrow\mathrm{SPT}_2
 \longrightarrow\mathrm{SPT}_3
 \longrightarrow\mathrm{SPT}_1.
 \label{gen:SPTtriality}
\end{equation}
Since none of these phases is invariant under translation, a unique gapped ground state cannot preserve both translation and the $\operatorname{Rep}(D_8)$ symmetry. This establishes the following LSM theorem.

\setcounter{theorem}{0}
\begin{theorem}
Any qubit chain with $L\in3\mathbb Z$ preserving the three-spin duality, i.e., $\kappa=\mu=1$, and translation symmetry cannot have a unique gapped ground state.
\end{theorem}

A representative model realizing these symmetries is the self-dual three-spin Ising chain,
\begin{equation}
 H=-\sum_j\left(
 Z_{j-1}Z_jZ_{j+1}+X_j
 \right).
 \label{gen:threeSpinIsing}
\end{equation}
Previous numerical studies found that this model has a central charge $c\simeq1$~\cite{kolb1986conformal,igloi1986series,PhysRevB.108.214430,alcaraz1987conformal}. Its gaplessness is enforced by the LSM obstruction derived above.

By contrast, combining this BPM duality with inversion symmetry does not give rise to an analogous LSM obstruction. Furthermore, for $N=2$ and $\mu=0$, neither translation nor inversion symmetry introduces an LSM obstruction. We demonstrate these results in the Supplemental Material by explicitly constructing symmetric Hamiltonians with unique gapped ground states.

\parR{Noninvertible LSM theorem in qudit chains}
We now extend the LSM constraints to qudit chains with $N>2$. Since $2\mu=0\pmod N$ and $\gcd(\kappa,N)=1$, we have $\kappa^{-1}\mu=\mu\pmod N$, and the kernel condition reduces to the recurrence relation $f_{j+1}=f_{j-1}-\mu f_j$. This relation can also be iterated backward as $f_{j-1}=f_{j+1}+\mu f_j$, so each solution is uniquely determined by two initial values $(f_1,f_0)\in(\mathbb Z_N)^2$.

To parameterize these solutions, we introduce two basis sequences $f_j^1$ and $f_j^2$ with initial values $f_1^1=1$, $f_0^1=0$ and $f_1^2=0$, $f_0^2=1$, respectively. All possible solutions can be parametrized by $g=(g_1,g_2)^{\mathsf T}\in(\mathbb Z_N)^2$ and take the form
\begin{equation}
 f_j(g)=g_1f_j^1+g_2f_j^2,\qquad
 f_1(g)=g_1,\quad f_0(g)=g_2.
 \label{gen:fdefinition}
\end{equation}
For system sizes compatible with periodic boundary conditions, these sequences define the modulated symmetry operators $U_g=\prod_j X_j^{f_j(g)}$. 

We now establish the following result.
\setcounter{theorem}{1}
\begin{theorem}
For compatible system sizes, any qudit chain with $N>2$ preserving the internal three-site BPM duality defined by Eq.~\eqref{gen:A} and either translation or site-centered inversion symmetry cannot have a unique gapped ground state.
\end{theorem}
\textit{Proof.}
We first examine how lattice symmetries act on the kernel symmetry. Since $TX_jT^{-1}=X_{j+1}$, translation maps $f_j(g)$ to $h_j=f_{j-1}(g)$, whose initial values satisfy
\[
 \begin{pmatrix}h_1\\h_0\end{pmatrix}
 =\begin{pmatrix}f_1(Rg)\\f_0(Rg)\end{pmatrix},
 \qquad
 R=\begin{pmatrix}0&1\\1&-\mu\end{pmatrix}.
\]
The uniqueness of solutions to the recurrence implies $h_j=f_j(Rg)$ for all $j$, yielding $TU_gT^{-1}=U_{Rg}$. Compatibility with PBC for all kernel sequences requires $R^L=\mathbbm{1}$. For $\mu=N/2$, this condition gives $L\in6\mathbb Z$ when $N\equiv2\pmod4$ and $L\in4\mathbb Z$ when $N\equiv0\pmod4$. For $\mu=0$, it gives $L\in2\mathbb Z$.

Similarly, site-centered inversion $I_0:j\mapsto-j$ acts on the kernel symmetry as
\begin{equation}
 I_0U_gI_0^{-1}=U_{Sg},
 \qquad
 S=\begin{pmatrix}1&-\mu\\0&1\end{pmatrix}.
 \label{gen:Ssite}
\end{equation}
Let us determine the resulting constraints on SPT phases protected by the $(\mathbb Z_N)^2$ kernel symmetry. These phases are classified by the projective representations of the edge symmetry operators \cite{Chen:2010zpc,Chen:2011pg,PhysRevB.80.155131,PhysRevB.85.075125},
\begin{equation}
V_g V_h (V_g)^{-1}(V_h)^{-1}
=\omega^{k g^{\mathsf T}Jh},
\qquad
J=\begin{pmatrix}0&1\\-1&0\end{pmatrix},
\label{gen:projective}
\end{equation}
where $V_g$ acts on the right-edge modes and $k\in H^2(\mathbb Z_N^2,U(1))=\mathbb Z_N$ labels the SPT index.

Under translation and inversion, the projective
representations transform as
\begin{equation}
\begin{split}
&V_{g,T} V_{h,T}V^{-1}_{g,T}V_{h,T}^{-1}
=\omega^{k g^{\mathsf T}R^{-\mathsf T}JR^{-1}h}=\omega^{-k g^{\mathsf T}Jh},\\
&V_{g,I_0} V_{h,I_0}V^{-1}_{g,I_0}V^{-1}_{h,I_0}
=\omega^{-k g^{\mathsf T}S^{-\mathsf T}JS^{-1}h}=\omega^{-k g^{\mathsf T}Jh}.
\label{gen:projective}
\end{split}
\end{equation}
Here
$V_{g,T}=V_{R^{-1}g}$ and $ V_{g,I_0}= W_{S^{-1}g}$
denote the transformed boundary symmetry operators,
where $W$ acts on the left-edge modes and carries
the opposite projective class $-k$.
Thus, both translation and inversion send the
SPT index $k$ to $-k$. A necessary condition for
an SPT phase to preserve either symmetry is therefore
$2k=0 \text{ (mod $N$)}$.
For odd $N$, the only possibility is $k=0$,
while for even $N$, the additional possibility
$k=N/2$ is allowed.
We now show that neither candidate is invariant
under the BPM duality.

We first consider the trivial SPT phase with $k=0$ realized by the Hamiltonian
\[
H_{k=0}=-\sum_{j}(X_j+\text{h.c.}).
\label{eq:H0}
\]
Under the BPM duality, it is mapped to
\[
\widehat{H}_{k=0}=-\sum_{j}(Z_{j-1}^{\kappa}Z_j^{\mu}
 Z_{j+1}^{-\kappa}+\text{h.c.}).
\]
The dual Hamiltonian has
 $N^2$ degenerate ground states, which are
$\ket{000\cdots}$ and the states
$\ket{s_j=f_j}$ for all nonzero kernel
vectors.
Thus, the BPM duality maps the trivial SPT
phase to an SSB phase,
excluding $k=0$ as a BPM-invariant SPT phase.

For even $N$, we can construct the SPT Hamiltonian with $k=N/2$ \cite{Supple}
\begin{equation}
H_{N/2}=-\sum_{j=1}^L(i^\epsilon X_jB_j^{N/2}+\text{h.c.}).
 \label{eq:Hq}
\end{equation}
Here $B_j$ is defined in Eq.~\eqref{gen:intertwine}
and $\epsilon=\mu \text{ (mod 2)}\in\{0,1\}$. The BPM duality maps this Hamiltonian to
\begin{equation}
 \widehat H_{N/2}=-\sum_j(i^\epsilon B_jX_j^{N/2}+\text{h.c.}).
 \label{eq:dualHq}
\end{equation}
Since all terms mutually commute,
the ground states satisfy
$i^\epsilon B_jX_j^{N/2}=1$.
Squaring this relation gives
$(i^\epsilon B_jX_j^{N/2})^2
 =Z_{j-1}^{2\kappa}Z_{j+1}^{-2\kappa}.
 \label{eq:dualsquare}$
Therefore $
 Z_{j-1}^2=Z_{j+1}^2$ in the ground-state
subspace, implying long-range order
$\langle Z^2_{2i-1} Z^{-2}_{2j-1}\rangle=\langle Z^2_{2i} Z^{-2}_{2j}\rangle=1$. These order parameters  break the kernel operators $U_{(1,0)}$ and  $U_{(0,1)}$, which implies that the dual Hamiltonian realizes
a kernel-SSB phase and excludes $k=N/2$
as a BPM-invariant SPT phase. This completes the proof \footnote{Indeed the BPM induces a one-to-one correspondence between the phases with kernel symmetries as shown in \cite{li2026generalized}.}.

Moreover, when neither translation nor site-centered inversion is imposed, the BPM duality symmetry allows a unique gapped ground state with kernel-SPT index $k=-\kappa$ \cite{Suppl}. Hence this LSM constraint arises from the interplay between internal and crystalline symmetries.

\begin{figure}[t]
\centering
\includegraphics[width=0.5\textwidth]{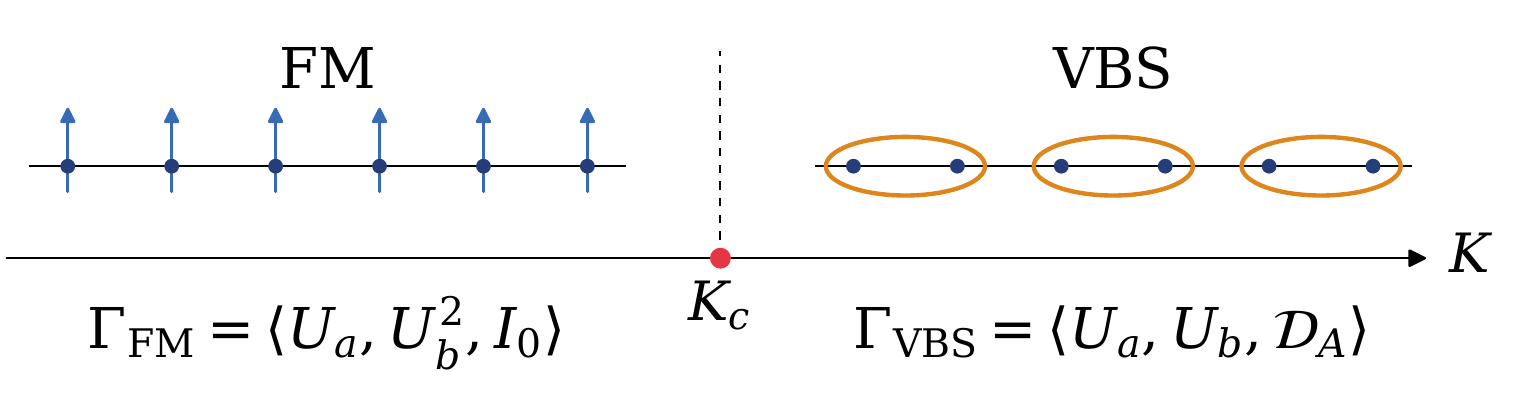}
\caption{Schematic phase diagram for the model in Eq.~\eqref{cone:Hqudit}. $\Gamma$ denotes the generator of unbroken symmetry.}
\label{fig:phase diagram}
\end{figure}

\begin{figure*}[t]
\centering
\includegraphics[width=0.8\textwidth]{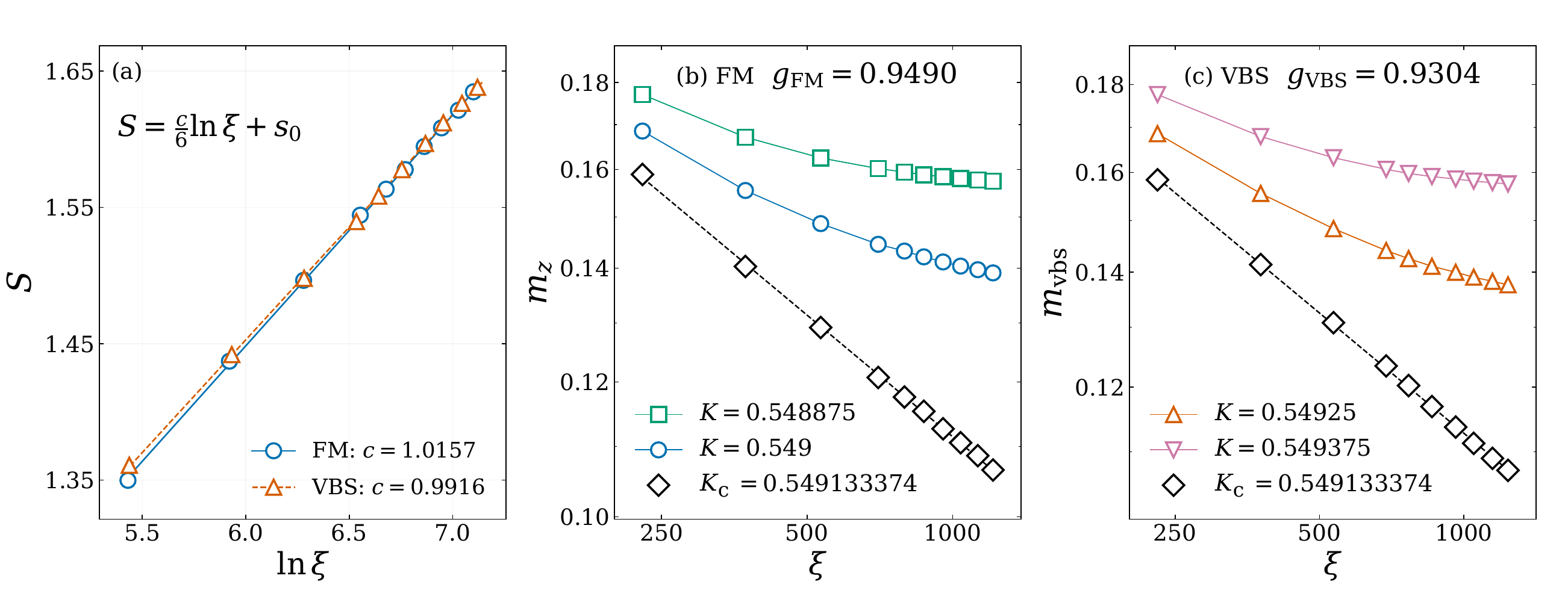}
\caption{(a) Entanglement entropy $S$ vs the correlation length $\xi$ at the DQCP, computed using two iMPS ansätze that preserve unbroken symmetries of the FM and VBS phases. (b,c) Log–log plots of the FM and VBS order parameters as functions of $\xi$. Power-law fits at $K_c$ yield the scaling dimensions $g/4$.}
\label{fig:order parameter}
\end{figure*}

\begin{figure*}[t]
\centering
\includegraphics[
  width=0.85\linewidth
]{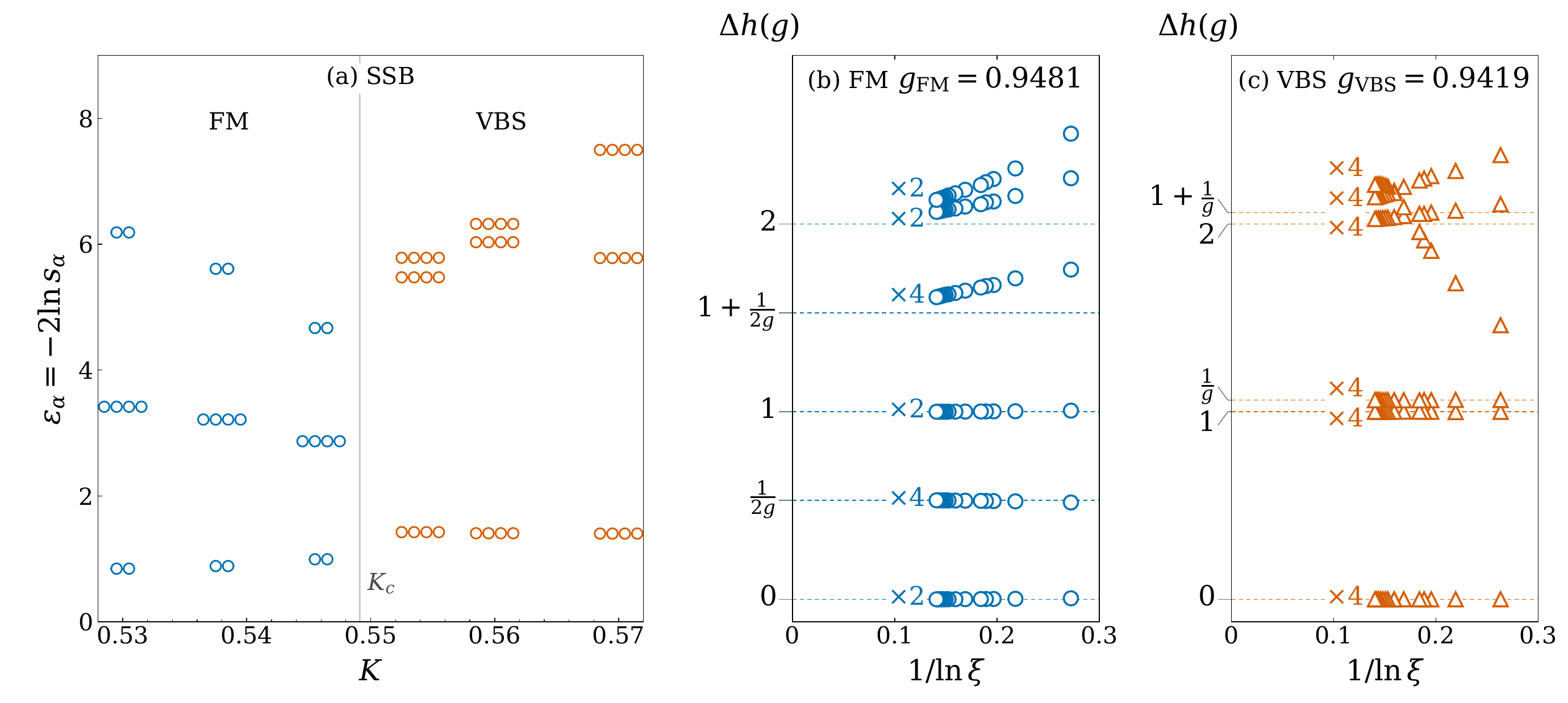}
\caption{(a) Entanglement spectra of SSB phases.  (b-c) Entanglement spectra of DQCP for two iMPS ansätze.  Each spectrum is shifted so that the lowest level lies at $\Delta h=0$ and rescaled using the reference level at $\Delta h=1$. The Luttinger parameter $g$ is extracted from the levels at $\Delta h=1/(2g)$ and $ \Delta h=1/g$.}
\label{fig:EE}
\end{figure*}
\parR{DQCP with noninvertible and inversion symmetry }
As an application of the LSM theorem, we construct an $N=4$ qudit model that preserves the internal BPM symmetries together with site-centered inversion, while explicitly breaking translation. This model realizes a DQCP between phases that spontaneously break the internal noninvertible symmetry and inversion symmetry, respectively. 

The microscopic Hamiltonian is
\begin{align}
 H={}&\frac{\Lambda}{2}\sum_j(1-C^2_j)
 -J_d\sum_n\bigl(\mathcal X^{(2)}_{4n-1}+\mathcal Z^{(2)}_{4n-1}\bigr)
 \nonumber\\[-2pt]
 &-\sum_r\bigl(J_x\mathcal X^{(2)}_{2r}+J_z\mathcal Z^{(2)}_{2r}\bigr)
 +K\sum_r\bigl(\mathcal X^{(4)}_{2r}+\mathcal Z^{(4)}_{2r}\bigr)\nonumber
 \\&+\eta\sum_j M_jM_{j+1},
 \label{cone:Hqudit}
\end{align}
where $
 C_j=Z_{j-1}X_jZ_{j+1},~
 M_j=(C_j+C_j^\dagger)/2$, and
\begin{equation}
 \begin{aligned}
 \mathcal X^{(d)}_{j}
 &=\frac12 M_jM_{j+d}
 \left[
 1+\left(\prod_{a=1}^{d-1}C_{j+a}\right)^2
 \right],\\
 \mathcal Z^{(d)}_{j}
 &=\frac12\left[
 (Z_jZ_{j+d})^2+
 \left(\prod_{a=1}^{d/2}X_{j+2a-1}\right)^2
 \right]
 \end{aligned}
\end{equation}
define the distance-$2+d$ and distance-$d$ interactions, with
$d=2,4$.

The Hamiltonian preserves the BPM duality with $\kappa=1$ and $\mu=2$  and  the kernel symmetries generated by $U_a=\prod_nX_{4n+1}X_{4n+2}^{2}X_{4n+3},~ U_b=\prod_nX_{4n+2}X_{4n+3}^{2}X_{4n+4}$ and site-centered inversion.

We perform
numerical calculations using infinite MPS (iMPS) ~\cite{PhysRevLett.69.2863,mcculloch2008infinite,PhysRevLett.98.070201,PhysRevLett.102.255701} fixing
$\Lambda=6$, $J_d=1$, $J_x=0.5$ and $J_z=1.5$ and $\eta=0.05$ \cite{Suppm}. The results are shown in
Fig.~\ref{fig:phase diagram} and Fig.~\ref{fig:order parameter}.
For $K<K_c$, the system is in the FM phase, which spontaneously breaks an internal
symmetry while preserving inversion.
For $K>K_c$, the system enters the VBS phase, which preserves the internal symmetry
but spontaneously breaks inversion.

The two phases are characterized by the order parameters
\begin{equation} \label{cone:quditorderparameters}
\begin{split}
 m_z&=\frac{1}{L}\sum_j Z_{2j}^2,\\
 m_{\rm VBS}&=\frac{1}{L}\sum_j(-1)^j
 \bigl(\mathcal X^{(2)}_{2j}+\mathcal Z^{(2)}_{2j}-\mathcal X^{(2)}_{2j}\mathcal Z^{(2)}_{2j}\bigr).
 \end{split}
\end{equation}
The onsite FM order parameter is odd under $U_b$ and  breaks $D_A$
whereas the VBS order parameter, which acts on next-nearest-neighbour qudits, only breaks inversion.

For the critical point,
our numerical results support a compact-boson
conformal field theory:
\begin{align}
 H_{\text{CFT}}&=\frac{v}{2\pi}\int dx\,
 \left[\frac{g}{4}(\partial_x\phi)^2+g^{-1}(\partial_x\theta)^2\right],
 \label{cone:dimensions}
\end{align}
with compactification radius $2\pi$.
The two order parameters correspond to
$m_z\sim\sin\theta,~
m_{\text{VBS}}\sim\cos\theta$ with scaling dimension $g/4$. We further show that this DQCP is stable under symmetric perturbations in the End Matter.

\parR{Entanglement spectrum}
We further characterize the entanglement spectra of the SSB phases and the DQCP using iMPS \cite{lauchli2013operator,RevModPhys.93.045003} as shown in Fig.~\ref{fig:EE}, which provides an intuitive picture of the transition. First, as discussed above, the VBS ground states preserve the internal symmetry and realize kernel SPT phases with indices $k=1,3$, which are exchanged by $I_0$, since the indices $k=0,2$ are incompatible with BPM duality. This SPT structure protects the fourfold entanglement degeneracy of the VBS phase. Upon entering the FM phase, inversion symmetry is restored while $U_b$ is spontaneously broken, leaving the internal symmetry subgroup $\Z_4^a\times\Z_2^{b^2}$ unbroken. SPT order protected by this subgroup has a $\Z_2$ classification. When restricted to this subgroup, the kernel SPT orders of VBS ground states both correspond to the same nontrivial SPT class. Thus, the SPT order protected by the common unbroken subgroup persists across the transition, accounting for the twofold entanglement degeneracy in the FM phase.

At the DQCP, iMPS ansätze preserving the unbroken symmetries of the FM and VBS phases yield distinct entanglement spectra, consistent with boundary conformal field theory spectra under the corresponding boundary conditions \cite{PhysRevLett.132.086503}. The respective symmetry-protected degeneracies persist in both ansätze, providing evidence for the ``gapless SPT'' order \cite{PhysRevX.11.041059,PhysRevLett.129.210601,PhysRevB.104.075132,SciPostPhys.17.1.013,PhysRevB.107.245127,SciPostPhys.18.5.153,PhysRevB.111.115161,96gp-fq3j,YU20261} of the critical point (see the End Matter).

\parR{Conclusion and discussion}
We have presented new LSM theorems for qubit and qudit chains arising from nontrivial algebraic relations between internal noninvertible symmetries and lattice translation or inversion symmetries. As an illustrative example, we construct a $\Z_4$ qudit model that realizes a stable DQCP between two distinct symmetry-breaking phases. We further characterize the critical point through its nontrivial entanglement-spectrum structure. Extending these results to systems involving more general magnetic space groups is an interesting direction for future work.

Recent work has also explored DQCPs between distinct phases with spontaneously broken noninvertible symmetry, but there the noninvertible symmetry is intertwined with lattice translation and does not impose an LSM constraint~\cite{8tsg-q9j6}. 

\parR{Acknowledgments.}
L.H.L. greatly acknowledges Ruizhen Huang for the discussion on iMPS. L.H.L. and Z.B. acknowledge a startup fund and a Quantum SuperSEED fund (ICDS\_QS25\_029093) from the Institute for Computational and Data Sciences at the Pennsylvania State University.  Z.B. also acknowledges support from NSF under Award No.~DMR-2339319. We thank OpenAI for providing access through the ChatGPT for Academic Researchers program and acknowledge the use of its tools for scientific discussions and manuscript editing. 
\bibliography{bib}

\newpage

\section{End Matter}

\parR{Low energy effective model for the DQCP }
Although the qudit Hamiltonian in Eq.~\eqref{cone:Hqudit} appears complicated, its low-energy physics becomes transparent through a mapping to an effective qubit chain. Since every term in Eq.~\eqref{cone:Hqudit} commutes with every $C_j^2$, sufficiently large $\Lambda$ confines the low-energy states to the sector with $C_j^2=+1$ for all $j$~\cite{Append}. Applying the entangling unitary $\Ucl=\prod_j\mathrm{CZ}^{(4)}_{j,j+1}$, where $\mathrm{CZ}^{(4)}\ket{s,t}=i^{st}\ket{s,t}$, gives $\Ucl^\dagger C_j^2\Ucl=X_j^2$. The local $X_j^2=+1$ subspace is two-dimensional and can therefore be identified with a qubit through $\frac{\ket{0}_j+\ket{2}_j}{\sqrt2}\longleftrightarrow\ket{\uparrow}_j$ and $\frac{\ket{1}_j+\ket{3}_j}{\sqrt2}\longleftrightarrow\ket{\downarrow}_j$.

Projecting Eq.~\eqref{cone:Hqudit} onto this
subspace yields the effective qubit Hamiltonian
\begin{align}
 H_{\rm q}
 ={}&-J_d\sum_n
 \bigl(\sigma^x_{4n-1}\sigma^x_{4n+1}
       +\sigma^z_{4n-1}\sigma^z_{4n+1}\bigr)
 \nonumber\\
 &-\sum_r\bigl(J_x\sigma^x_{2r}\sigma^x_{2r+2}
               +J_z\sigma^z_{2r}\sigma^z_{2r+2}\bigr)
 \nonumber\\
 &+K\sum_r\bigl(\sigma^x_{2r}\sigma^x_{2r+4}
               +\sigma^z_{2r}\sigma^z_{2r+4}\bigr)
 +\eta\sum_j\sigma^x_j\sigma^x_{j+1}.
 \label{cone:Hqubit}
\end{align}
The kernel symmetries reduce to
\begin{align}
 U_a^{\rm eff}=
 \prod_{k=1}^{L/2}\sigma_{2k-1}^y\sigma_{2k}^z,~U_b^{\rm eff}
=\prod_{k=1}^{L/2}\sigma_{2k-1}^z\sigma_{2k}^y.
\end{align}
Site-centered inversion retains its conventional
spatial action in the effective qubit chain.
The BPM duality reduces to
\begin{equation}
 D_{\rm eff}
 =Q(1+U_a^{\rm eff})(1+U_b^{\rm eff}), ~ Q=\prod_j\sigma_j^z.
\end{equation}

At $\eta=0$, the odd and even sublattices decouple.
The odd sublattice forms independent dimers with a
unique gapped ground state, while the even sublattice
realizes the anisotropic frustrated FM--VBS spin chain
studied in Refs.~\cite{PhysRevB.99.165143,PhysRevB.99.075103}.
For $J_x=0.5$, $J_z=1.5$, and $K\simeq0.55$,
the even chain undergoes a continuous FM--VBS
transition.

In the effective qubit description, the qudit
order parameters reduce to the conventional
$z$-ferromagnetic and VBS order parameters
on the even sublattice:
\begin{equation}
\label{cone:qubitorderparameters}
 m_z=\frac{1}{L}\sum_j\sigma^z_{2j},~
 m_{\rm VBS}=\frac{1}{L}\sum_{j}(-1)^j
 (\vec{\sigma}_{2j}\cdot \vec{\sigma}_{2j+2}).
\end{equation}
The FM order parameter is odd under the internal
symmetry $U_b^{\rm eff}$, whereas the VBS order
parameter is odd under inversion $I_0$.
Hence the two FM ground states
$\ket{F_\pm}$ are exchanged by $U_b^{\rm eff}$, while
the two VBS ground states $\ket{V_{A/B}}$
are exchanged by $I_0$.

The two phases also exhibit distinct actions
of the noninvertible BPM symmetry. Since $Q$ commutes with both order parameters and the qubit Hamiltonian, it acts trivially on SSB ground states. Then we obtain
\begin{equation}
\begin{aligned}
D_{\rm eff}\ket{F_\pm}=2\bigl(\ket{F_+}+\ket{F_-}\bigr),\quad D_{\rm eff}\ket{V_{A/B}}=4\ket{V_{A/B}}.
\end{aligned}
\end{equation}
Thus, the duality maps the two FM ground states
to the same equal-weight superposition, whereas
each VBS ground state is an eigenstate of $D_A$.

\parR{Entanglement features of the DQCP}
As discussed in Ref.~\cite{PhysRevLett.132.086503}, the entanglement
spectrum of an iMPS with finite bond dimension corresponds to  the energy spectrum of a boundary conformal field theory with a
physical boundary condition $B_p$ and an entanglement boundary
condition $B_e$.

The entanglement boundary condition is determined by the microscopic
regularization near the entanglement cut. As shown in the Supplemental Material, the
reduced density matrix admits the block decomposition
\[
\rho \simeq \mathbb{I}_2\otimes(R_0\oplus R_1).
\]
Here $\simeq$ denotes equality up to a local unitary
transformation, which leaves the
entanglement spectrum unchanged. The factor $\mathbb{I}_2$
guarantees at least a twofold degeneracy of every
entanglement level. The blocks $R_0$ and $R_1$ correspond to
the eigenvalues $+1$ and $-1$ of $Z_{2i}^2$, respectively,
with the entanglement cut located between sites $2i$ and
$2i+1$.


Since $
Z^2_{2i}\sim \sin\theta$ in the continuum limit, the corresponding entanglement boundary
conditions flow to
\begin{equation}
B_{e,0}\longrightarrow D_\theta(\pi/2),
\qquad
B_{e,1}\longrightarrow D_\theta(-\pi/2).
\label{eq:RGassignment}
\end{equation}
Here $D_\theta(\delta)$ denotes the Dirichlet boundary condition
$\theta=\delta$.

The physical boundary condition, by contrast, is determined by the
effective boundary perturbation induced by the finite bond
dimension. We consider MPS that preserve the unbroken symmetries of
the neighboring FM or VBS phase. The corresponding physical boundary
conditions are
\begin{equation}
\begin{aligned}
B_p^{\rm FM}&=D_\theta(\pi/2)\ \text{or}\ D_\theta(-\pi/2),\\
B_p^{\rm VBS}&=D_\theta(0)\ \text{or}\ D_\theta(\pi).
\end{aligned}
\label{eq:Bp}
\end{equation}

For Dirichlet boundary conditions at both ends, the conformal levels
are
\begin{equation}
h_{w,l}
=
\frac{2}{g}
\left(
w+\frac{\theta_e-\theta_p}{2\pi}
\right)^2+l,
~
w\in\mathbb Z,~ l=0,1,\ldots,
\label{eq:DD}
\end{equation}
where $w$ labels the winding sector and $l$ is the oscillator level.

We first consider the FM physical boundary,
$B_p^{\rm FM}$, for which $\theta_p=\pm\pi/2$. Combining the two
entanglement-boundary sectors with the additional twofold copy in
$\rho$, we obtain the following low-lying entanglement levels.
For $1/2<g<1$, the first five distinct levels and their total
multiplicities are
\begin{center}
\begin{tabular}{|c|c|c|}
\hline
$h$ & Multiplicity & $(w,l, \theta_e-\theta_p)$\\
\hline
$0$ & $2$ & $(0,0,0)$\\
\hline
$1/(2g)$ & $4$ & $(0,0,\pm\pi),$ $(\mp1,0,\pm\pi)$\\
\hline
$1$ & $2$ & $(0,1,0)$\\
\hline
$1+1/(2g)$ & $4$ & $(0,1,\pm \pi),$ $(\mp 1,1,\pm\pi)$\\
\hline
$2$ & $4$ & $(0,2,0)$\\
\hline
\end{tabular}
\end{center}
Here the sign $\pm$ corresponds to the choice $\theta_p=-\pi/2$ or $\theta_p=\pi/2$.
At oscillator level $l=2$ with $h=2$, the two descendant states, $a_{-2}\ket{w=0}$ and $a_{-1}^2\ket{w=0}$, are split by finite-bond-dimension effects as shown in Fig.~\ref{fig:EE}, where $\Delta h=h$.

We next consider the VBS physical boundary,
$B_p^{\rm VBS}$, with $\theta_p=0$ or $\pi$. The two entanglement
boundary sectors correspond to shifts $(\theta_e-\theta_p)/2\pi=1/4$ and $(\theta_e-\theta_p)/2\pi=-1/4$ in
Eq.~\eqref{eq:DD}. Their spectra are identical upon reversing the
winding number in one sector. Together with the additional twofold
copy in $\rho$, this implies that every BCFT level has a
multiplicity of at least four. For $1/2<g<1$, the first five distinct conformal levels and their
total multiplicities are
\begin{center}
\begin{tabular}{|c|c|c|}
\hline
$h$ & Multiplicity & $(w,l)$\\
\hline
$h_0=1/(8g)$ & $4$ & $(0,0)$\\
\hline
$h_0+1$ & $4$ & $(0,1)$\\
\hline
$h_0+1/g$ & $4$ & $(1,0)$\\
\hline
$h_0+2$ & $8$ & $(0,2)$\\
\hline
$h_0+1+1/g$ & $4$ & $(1,1)$\\
\hline
\end{tabular}
\end{center}
Here the winding labels refer to the branch with shift $-1/4$.
 At oscillator level $l=2$ with $h=h_0+2$, the two descendant states $a_{-2}\ket{w=0}$ and $a_{-1}^2\ket{w=0}$ are split by finite-bond-dimension effects as shown in Fig.~\ref{fig:EE}, where $\Delta h=h-h_0$.

\parR{Stability of DQCP}Let us analyze the stability of the DQCP
under symmetric perturbations constructed from the vertex operators 
 $V_{m,n}= e^{i m\theta+i n\phi}$ with scaling dimension $
 \Delta_{m,n}=\frac{gm^2}{4}+\frac{n^2}{g}$.
The symmetry actions on the continuum fields are
\begin{align}
U_a:&\quad (\theta,\phi)\mapsto(\theta,\phi+\pi),~U_b:\quad (\theta,\phi)\mapsto(-\theta,-\phi+\pi)\nonumber\\
 I_0:&\quad \theta(x)\mapsto\pi-\theta(-x),
 \quad \phi(x)\mapsto\phi(-x).
 \label{cone:fieldactions}
\end{align}

The $U_a$ symmetry requires the magnetic
charge $n$ to be even. Consequently, all vertices
with nonzero magnetic charge satisfy
\begin{equation}
 |n|\ge2,\qquad \Delta_{m,n}\ge\frac4g>2
 \quad\text{for }g<2.
 \label{cone:magneticbound}
\end{equation}
Such perturbations are therefore irrelevant
throughout this regime.

For $n=0$, the combined constraints of
$U_b$ and $I_0$ restrict the allowed
electric perturbations to
\begin{equation}
 \cos(2p\theta),\qquad \Delta=p^2g,\qquad p=1,2,\ldots.
 \label{cone:electricfamily}
\end{equation}
In the interval $\frac12<g<2$, the only
relevant primary perturbation is
$\lambda\cos(2\theta)$, whose sign selects
the FM or VBS phase.
The transition therefore requires tuning
only one relevant coupling. In the Supplemental Material, we further analyze the stability against descendants of primary fields.

\parR{Gapless SPT order}  The ``gapless SPT'' order at the critical point is protected by $U_a$ and $U_b^2$, which remain unbroken throughout the entire phase diagram.  Its edge-mode signature can be established directly from the exact constraint $
C_j^2=Z_{j-1}^2X_j^2Z_{j+1}^2=1$,
which is satisfied by the ground state(s). For an open chain of even length $L$,  $U^2_b$ factorizes into $Z^2_1$ and $Z^2_{L-1}X^2_L$ on two boundaries for ground states.
Each local term \(h_j\) that preserves \(U_b^2\) and commutes with $C^2_i$ therefore satisfies
\[
\big([h_j,Z_1^2]\,Z_{L-1}^2X_L^2
+Z_1^2[h_j,Z_{L-1}^2X_L^2]\big)|_{\text{G.S.}}=0
\]
within the ground-state subspace.
Then locality ensures that \(h_j\) overlaps with at most one of the two boundary operators. Hence, at least one of the two commutators on the left-hand side vanishes. Thus we obtain
\[
[h_j,Z_1^2]|_{\text{G.S.}}=[h_j,Z_{L-1}^2X_L^2]|_{\text{G.S.}}=0.
\]
Therefore the ground states can be chosen as eigenstates of \(Z_1^2\) and \(Z_{L-1}^2X_L^2\).  Since these boundary symmetry operators anticommute with $U_a$, there are at least two exactly degenerate ground states associated with edge modes.





\clearpage
\appendix 
\onecolumngrid 
\color{black}

 \begin{center}
	\Large{\textbf{Supplemental Materials for ``Noninvertible Lieb–Schultz–Mattis Constraints and Deconfined Criticality"}}
\end{center}

The sections of the SM and their purposes are arranged as follows:

\begin{enumerate}
    \item Section~A uses the Kennedy--Tasaki transformation to demonstrate the permutation of the three $\operatorname{Rep}(D_8)$ SPT phases under translation.

    \item Section~B discusses the remaining $N=2$ cases without an LSM constraint.

    \item Section~C provides a detailed analysis of the SPT phase protected by the kernel symmetry for $k=N/2$ and characterizes the corresponding symmetry-breaking phase obtained under the BPM duality.

    \item Section~D constructs Hamiltonians with a unique gapped ground state that preserve both the internal kernel symmetry and the duality symmetry, demonstrating the absence of an LSM-type constraint from these internal symmetries alone.

    \item Section~E derives the correspondence between operators in the original qudit chain and those in the effective qubit chain, and determines the symmetry actions in the low-energy field theory.

    \item Section~F analyzes the stability of the DQCP against spatial-derivative perturbations arising as descendants of primary fields.

    \item Section~G establishes the correspondence between the entanglement spectrum and the energy spectrum of the associated boundary conformal field theory.

    \item Section H presents additional numerical results for the order parameters in the regime $0.53<K<0.57$ at $\eta=0.05$, together with critical properties of the DQCP at $\eta=0.1$ and $\eta=0.15$.
\end{enumerate}

\section{\large A: The three-cycle of \texorpdfstring{$\operatorname{Rep}(D_8)$}{Rep(D8)} SPT phases}
\label{app:kttriality}

Using the Kennedy--Tasaki (KT) transformation, we show that
the automorphism
\begin{equation}
 \Phi:\quad
 U_1\mapsto U_2,\qquad
 U_2\mapsto U_1U_2,\qquad
 D\mapsto D
 \label{eq:trans}
\end{equation}
cyclically permutes the three $\operatorname{Rep}(D_8)$
SPT phases. Here $U_1$ and $U_2$ are the invertible
generators, while $D$ is the noninvertible generator.
This establishes the translation-induced triality
of the BPM-self-dual SPT phases for $N=2$ and
$\kappa=\mu=1$, as asserted in the main text.

The KT transformation, understood as twisted gauging,
relates the $\operatorname{Rep}(D_8)$ categorical symmetry
to an anomalous $(\mathbb Z_2)^3$ group symmetry.
In the KT-dual description, $U_1$ and $U_2$ correspond
to the generators $\widehat U_1$ and $\widehat U_2$,
respectively, while the noninvertible sector gives rise
to a third invertible generator $\widehat U_3$.
The resulting symmetry carries a type-III mixed anomaly
represented by the 3-cocycle
\begin{equation}
 \omega(g,h,k)=(-1)^{g_1h_2k_3},
 \qquad g,h,k\in(\mathbb Z_2)^3,
 \label{tri:cocycle}
\end{equation}
where group elements are parameterized as
$g=(\widehat U_1)^{g_1}(\widehat U_2)^{g_2}
(\widehat U_3)^{g_3}$, with $g_i\in\{0,1\}$.

Under the KT transformation, the three
$\operatorname{Rep}(D_8)$ SPT phases map to distinct
symmetry-breaking phases of the dual theory,
distinguished by their unbroken subgroups \cite{PhysRevLett.133.116601}:
\begin{equation}
 \begin{array}{c|c}
 \operatorname{Rep}(D_8)\text{ SPT}
 & \text{Unbroken subgroup after KT}\\ \hline
 \mathrm{SPT}_1&H_0=\langle\widehat U_3\rangle\\
 \mathrm{SPT}_2&H_1=\langle\widehat U_3\widehat U_2\rangle\\
 \mathrm{SPT}_3&H_2=\langle\widehat U_3\widehat U_1\rangle
 \end{array}
 \label{tri:phasetable}
\end{equation}

We denote the automorphism induced by
Eq.~\eqref{eq:trans} in the KT-dual description
by $\widehat\Phi$. Its action on the first two
generators is
\begin{equation}
 \widehat\Phi(\widehat U_1)=\widehat U_2,\qquad
 \widehat\Phi(\widehat U_2)=\widehat U_1\widehat U_2.
 \label{tri:dualPhi12}
\end{equation}
To determine its action on the remaining generator,
we write
\begin{equation}
 \widehat\Phi(\widehat U_3)
 =\widehat U_1^x\widehat U_2^y\widehat U_3^z,
 \qquad x,y,z\in\mathbb Z_2.
 \label{tri:dualPhi3ansatz}
\end{equation}
The dual automorphism must satisfy two consistency
conditions:
\begin{equation}
 \begin{aligned}
 \widehat\Phi^3(\widehat U_3)=\widehat U_3,~
 \widehat\Phi(\widehat U_1\widehat U_2\widehat U_3)=\widehat U_1\widehat U_2\widehat U_3.
 \end{aligned}
 \label{tri:consistency}
\end{equation}
The first follows from the order-three action of $\Phi$.
The second follows from preservation of the anomaly class:
the subgroup generated by
$\widehat U_1\widehat U_2\widehat U_3$
is the unique anomalous $\mathbb Z_2$ subgroup.
Indeed, the restriction of Eq.~\eqref{tri:cocycle}
to a subgroup generated by $a=(a_1,a_2,a_3)$
is nontrivial precisely when $a_1a_2a_3=1$.

In terms of $x,y,z$, the first condition requires
$z=1$ but places no further restriction on $x,y$.
Using Eq.~\eqref{tri:dualPhi12}, the second condition becomes
\begin{equation}
 \widehat U_1^{1+x}\widehat U_2^y\widehat U_3^z
 =\widehat U_1\widehat U_2\widehat U_3,
\end{equation}
which uniquely fixes $x=0$ and $y=z=1$.
Consequently,
\begin{equation}
 \widehat\Phi(\widehat U_3)=\widehat U_2\widehat U_3.
 \label{tri:dualPhi3}
\end{equation}

These transformation rules give
\begin{equation}
 \begin{aligned}
 \widehat\Phi(H_0)
 &=\langle\widehat U_2\widehat U_3\rangle=H_1,\\
 \widehat\Phi(H_1)
 &=\langle\widehat U_1\widehat U_3\rangle=H_2,\\
 \widehat\Phi(H_2)
 &=\langle\widehat U_3\rangle=H_0.
 \end{aligned}
 \label{tri:subgroupcycle}
\end{equation}
Thus, $\widehat\Phi$ cyclically permutes the three
unbroken subgroups. Through the KT correspondence
in Eq.~\eqref{tri:phasetable}, this establishes
the triality of the original SPT phases:
\begin{equation}
 \Phi:\quad
 \mathrm{SPT}_1\longrightarrow
 \mathrm{SPT}_2\longrightarrow
 \mathrm{SPT}_3\longrightarrow
 \mathrm{SPT}_1.
 \label{tri:SPTcycle}
\end{equation}
\\~\\

\section{\large B: The remaining cases for $N=2$}
We first consider the BPM duality with $N=2$, $\kappa=\mu=1$. In this case, site-centered inversion, by contrast, is an order-two transformation. Thus it acts as an involution on the set of three SPT phases and therefore leaves at least one phase fixed. Hence, this phase-permutation argument produces no mixed LSM obstruction involving inversion.

For $\mu=0$, the kernel symmetry is generated by independent spin flips on the even and odd sublattices. One-site translation exchanges these generators, whereas site-centered inversion preserves each of them. In this case, the absence of a mixed obstruction is demonstrated directly by the cluster Hamiltonian
\begin{equation}
H_{\rm cl}=-\sum_j Z_{j-1}X_jZ_{j+1}\qquad(N=2).
\label{gen}
\end{equation}
This Hamiltonian has a unique gapped ground state and preserves the kernel symmetry, BPM duality, one-site translation, and site-centered inversion.
\\~\\
\section{\large C: Kernel SPT phase with \texorpdfstring{$k=N/2$}{k=N/2} and the dual SSB phase}
In this appendix, we consider the model
\begin{equation}
 H=-\sum_j(i^\epsilon X_j
 B^q_j+\text{h.c.})=-\sum_j i^\epsilon X_j
 Z_{j-1}^qZ_j^{q\mu}Z_{j+1}^q+\text{h.c.},
 \label{eq:Hq}
\end{equation}
where
\begin{equation}
 q=\frac N2,\qquad \epsilon=\mu\bmod2\in\{0,1\}
 \label{eq:q}
\end{equation}
and $N$ is even. Since $X_j$ and $B_j$ respect kernel symmetry, this model also respects kernel symmetry. We prove that it realizes the SPT phase of the kernel symmetry of three-site BPM with level $k=N/2$.

We first introduce the diagonal circuit
\begin{equation}
 W_q|\{s_j\}\rangle
 =i^{\epsilon\sum_j s_j^2}
  \omega^{q\sum_j s_js_{j+1}}|\{s_j\}\rangle.
 \label{eq:W}
\end{equation}
The onsite factor $i^{s^2}$ is well defined on $\Z_N$ for even $N$.
The two-site factors are commuting nearest-neighbor controlled phases,
and the circuit has bounded depth. Direct conjugation gives
\begin{equation}
 W_qX_jW_q^\dagger=C_j.
 \label{eq:conjugation}
\end{equation}
As a result,
\begin{equation}
 H_q=W_qH_0W_q^\dagger,\qquad H_0=-\sum_j (X_j+\text{h.c.}),
 \label{eq:unique}
\end{equation}
which shows the ground state is gapped and unique.

For an open chain, we retain only the bulk terms:
\begin{equation}
 H_q^{\mathrm{open}}
 =-\sum_{j=2}^{L-1}(i^\epsilon X_j
 Z_{j-1}^qZ_j^{q\mu}Z_{j+1}^q+\text{h.c.}).
 \label{eq:openH}
\end{equation}
The kernel recurrence and $\mu\bmod2=\epsilon$ imply
\begin{equation}
 q\bigl[f_{j-1}(g)+f_{j+1}(g)+\epsilon f_j(g)\bigr]=0\pmod N.
 \label{eq:bulkcancel}
\end{equation}
It follows that all interior clock factors cancel in
$W_q^\dagger U_gW_q$. Up to a scalar $\zeta(g)$,
\begin{equation}
 W_q^\dagger U_gW_q
 =\zeta(g)
 \left[X_1^{f_1(g)}Z_1^{qf_0(g)}\right]
 \left[X_L^{f_L(g)}Z_L^{qf_{L+1}(g)}\right]
 \prod_{j=2}^{L-1}X_j^{f_j(g)}.
 \label{eq:edgefactor}
\end{equation}

Taking compatible $L\in 6\Z$ for $N=2 \text{ (mod 4)}$ or $L\in 4\Z$ for $N=0 \text{ (mod 4)}$, so $f_L(g)=g_2$ and $f_{L+1}(g)=g_1$,
the edge actions in the disentangled basis can be chosen as
\begin{equation}
 V_g^L=X_1^{g_1}Z_1^{qg_2},\qquad
 V_g^R=X_L^{g_2}Z_L^{qg_1}.
 \label{eq:edges}
\end{equation}
Therefore
\begin{equation}
 V_g^RV_h^R(V_g^R)^{-1}(V_h^R)^{-1}
 =\omega^{q(g_1h_2-g_2h_1)}.
 \label{eq:edgecomm}
\end{equation}
This proves the SPT index $k=q=N/2$.

We now examine the BPM-dual Hamiltonian
\begin{equation}
 \widehat H_q
 =-\sum_j\left(i^\epsilon B_jX_j^q+\mathrm{h.c.}\right).
 \label{eq:dualHq}
\end{equation}
As shown in the main text, its ground states satisfy
\begin{equation}
 Z_{j-1}^2=Z_{j+1}^2
 \qquad\text{on the ground space}.
 \label{eq:order}
\end{equation}
This constraint fixes
$Z_j^2$ independently on the two sublattices.
The ground-state manifold therefore decomposes
into sectors labeled by
\begin{equation}
 Z_{2r-1}^2=\omega^{2a},\qquad
 Z_{2r}^2=\omega^{2b},\qquad a,b\in\Z_{N/2}.
 \label{eq:sectors}
\end{equation}

We show that each sector contains exactly one
ground state. Within a fixed sector $(a,b)$,
the local qudit configurations can be written as
\begin{equation}
 s_j=r_j+Nt_j/2,\qquad t_j\in\Z_2,\qquad
 r_j=\begin{cases}a,&j\ \text{odd},\\b,&j\ \text{even}.\end{cases}
 \label{eq:qubits}
\end{equation}
Thus, each site retains a two-dimensional
Hilbert space, with $X_j^q$ flipping $t_j$.
Introducing the effective Pauli operators
$\tau_j^x=X_j^q$ and
$\tau_j^z\ket{t_j}=(-1)^{t_j}\ket{t_j}$,
we obtain
\begin{equation}
 \begin{aligned}
 \widehat H_q^{\mathrm{eff}}
 &=-\sum_j\left[
 i^\epsilon\omega^{\mu r_j}
 \tau_{j-1}^z(\tau_j^z)^\epsilon
 \tau_{j+1}^z\tau_j^x
 +\mathrm{H.c.}\right]\\
 &=
 \begin{cases}
 -2\sum_j\omega^{\mu r_j}
 \tau_{j-1}^z\tau_j^x\tau_{j+1}^z,
 &\epsilon=0,\\[2pt]
 2\sum_j\omega^{\mu r_j}
 \tau_{j-1}^z\tau_j^y\tau_{j+1}^z,
 &\epsilon=1.
 \end{cases}
 \end{aligned}
 \label{eq:qubitstabilizers}
\end{equation}
Here $\omega^{\mu r_j}=\pm1$ since
$\mu=0$ or $N/2$.
In either case, the effective Hamiltonian
is a commuting-stabilizer model, whose
stabilizers admit a
unique  ground state.
Thus, each sector $(a,b)$ contains
exactly one ground state, giving a total
ground-state degeneracy of $(N/2)^2$.

We finally determine the spontaneous symmetry breaking (SSB)
pattern. Since $\mu=0\bmod N/2$, the kernel
recurrence reduces modulo $N/2$ to
$f_{j+1}=f_{j-1}$.
The kernel symmetry therefore acts on
the ground-state sectors as
\begin{equation}
 U_g:(a,b)\longmapsto(a+g_1,b+g_2) \text{ (mod $N/2$)}.
 \label{eq:sectoraction}
\end{equation}
This action is transitive, and the subgroup
leaving an individual sector invariant is generated by $U_{(N/2,0)}$ and
$U_{(0,N/2)}$.
Hence, for $N>2$, the dual Hamiltonian
realizes an SSB phase with
$G=\mathbb Z_N\times\mathbb Z_N$ spontaneously
broken to $\Gamma\cong\mathbb Z_2\times\mathbb Z_2$.
The resulting degeneracy is
$|G/\Gamma|=(N/2)^2$, in agreement with the
explicit ground-state construction.
\\~\\

\section{\large D: No LSM constraint for the internal symmetry}
\label{gen:chargeconstruction}

In this appendix, we will prove that for every $N>2$, $\gcd(\kappa,N)=1$, and $2\mu=0\pmod N$, there is a  Hamiltonian on the qudit chain with
 a unique gapped ground state,  which preserves the internal BPM duality and kernel symmetries.\\

\prlsection{Case 1: a cluster ground state when $\mu=0$.}
For $\mu=0$, we construct a BPM-invariant cluster Hamiltonian using the entangling unitary
\begin{equation}
 \mathrm{CZ}\ket{\{s_j\}}
 =\omega^{-\kappa\sum_{j=1}^L(-1)^j s_js_{j+1}}
 \ket{\{s_j\}}.
 \label{cc:clusterphase}
\end{equation}
Conjugating the trivial paramagnet by this unitary yields
\begin{equation}
 \begin{aligned}
 H_0=-\sum_{j=1}^L
 \bigl(\mathrm{CZ}X_j\mathrm{CZ}^\dagger
 +\mathrm{H.c.}\bigr)=-\sum_{j=1}^L
 \bigl(X_j^{(-1)^j}Z_{j-1}^{\kappa}
 Z_{j+1}^{-\kappa}+\mathrm{H.c.}\bigr),
 \end{aligned}
 \label{cc:clusterH}
\end{equation}
which has a unique gapped ground state
\begin{equation}
 \ket{\text{cluster}}
 =\mathrm{CZ}\bigotimes_{j=1}^L\ket{+}_j,
 \qquad
 \ket{+}=\frac{1}{\sqrt N}\sum_{s=0}^{N-1}\ket{s}.
\end{equation}

For $\mu=0$, the kernel symmetries are generated by
$U_o=\prod_j X_{2j-1}$ and $U_e=\prod_j X_{2j}$,
both of which are preserved by $H_0$.
To establish BPM invariance, define
$B_j=Z_{j-1}^{\kappa}Z_{j+1}^{-\kappa}$.
The BPM relations
$DX_j=B_jD$ and $DB_j=X_jD$,
together with $[X_j,B_j]=0$, imply
\begin{equation}
 D X_j^{(-1)^j}B_j
 =X_j B_j^{(-1)^j}D.
 \label{cc:clusterduality}
\end{equation}
Consequently, the full Hamiltonian satisfies
\begin{equation}
 [H_0,D]=0,
\end{equation}
establishing its BPM invariance.

Moreover, for an open chain of even length \(L\), the bulk ground-state constraints
\[
X_j^{(-1)^j}Z_{j-1}^{\kappa}Z_{j+1}^{-\kappa}
|\psi\rangle=|\psi\rangle,\qquad 1<j<L,
\]
imply that the kernel symmetry operators factorize into boundary contributions:
\[
U_o\doteq
\bigl(X_1Z_2^\kappa\bigr)Z_L^{-\kappa},
\qquad
U_e\doteq
Z_1^{-\kappa}\bigl(Z_{L-1}^{\kappa}X_L\bigr),
\]
where \(\doteq\) denotes equality within the ground-state subspace. The right-edge symmetry operators can therefore be chosen as
$V_o^R=Z_L^{-\kappa},
\quad
V_e^R=Z_{L-1}^{\kappa}X_L$. Then we obtain
\[
V_o^R V_e^R=\omega^{-\kappa}V_e^R V_o^R.
\]
Since \(U_o=U_{(1,0)}\) and \(U_e=U_{(0,1)}\) in the notation of the main text, this projective commutation relation identifies the kernel SPT index as \(k=-\kappa\pmod N\).
\\

\prlsection{Case 2: a local coordinate shear for $\mu=N/2$.}
For even $N$ and nonzero $\mu$, we choose
\begin{equation}
 p=
 \begin{cases}
 2, & 4\mid N,\\
 4, & N\equiv 2\pmod 4,
 \end{cases}
 \qquad
 L=(p+2)r,\qquad r\geq 2.
 \label{cc:blocksize}
\end{equation}
We partition the chain into $r$ unit cells, each containing
$p+2$ consecutive sites. The $j$-th cell consists of
a left endpoint $l_j=(p+2)(j-1)+1$,
$p$ internal sites
$I_j=\{l_j+1,\ldots,l_j+p\}$,
and a right endpoint $r_j=l_j+p+1$.
Writing
\begin{equation}
 \bm s=(s_1,\ldots,s_L)^{\mathsf T},
 \qquad
 \bm s_{I_j}
 =(s_{l_j+1},\ldots,s_{l_j+p})^{\mathsf T},
\end{equation}
the action of the BPM matrix $A$ on the internal sites
takes the form
\begin{equation}
 (A\bm s)_{I_j}
 =\mathsf B_p\bm s_{I_j}
 +E_p
 \begin{pmatrix}
 s_{l_j}\\s_{r_j}
 \end{pmatrix},
\end{equation}
where
\begin{equation}
 \mathsf B_p=
 \begin{pmatrix}
 \mu&-\kappa&&0\\
 \kappa&\mu&\ddots&\\
 &\ddots&\ddots&-\kappa\\
 0&&\kappa&\mu
 \end{pmatrix},
 \qquad
 E_p=(\kappa e_1,-\kappa e_p).
 \label{cc:pivotmatrix}
\end{equation}
Here $\mathsf B_p$ is the $p\times p$ matrix
acting on the internal sites, while $E_p$ encodes
their couplings to the two endpoints, with $e_a$
denoting the $a$-th standard basis vector.

In the basis ordered by the internal sites followed by
the two endpoints, the intra-cell part of $A$ reads
\begin{equation}
 A_{\rm cell}=
 \begin{pmatrix}
 \mathsf B_p&E_p\\
 -E_p^{\mathsf T}&A_{\rm end}
 \end{pmatrix},
 \qquad
 A_{\rm end}=
 \begin{pmatrix}
 \mu&0\\
 0&\mu
 \end{pmatrix}.
\end{equation}
For the above choice of $p$, the matrix $\mathsf B_p$
is invertible over $\mathbb Z_N$ and satisfies
\begin{equation}
 E_p^{\mathsf T}\mathsf B_p^{-1}E_p
 =
 \begin{pmatrix}
 \mu&-\kappa\\
 \kappa&\mu
 \end{pmatrix}.
 \label{cc:endpointidentity}
\end{equation}

We now perform a local change of variables within
each unit cell:
\begin{equation}
 \begin{aligned}
 \bm s_{I_j}
 =\bm u_{I_j}
 -\mathsf B_p^{-1}E_p
 \begin{pmatrix}
 v_{l_j}\\v_{r_j}
 \end{pmatrix}, s_{l_j}&=v_{l_j},~
 s_{r_j}=v_{r_j}.
 \end{aligned}
 \label{cc:localshear}
\end{equation}
This transformation is represented by the shear matrix
\begin{equation}
 \mathsf M_j=
 \begin{pmatrix}
 \mathbb I_p&-\mathsf B_p^{-1}E_p\\
 0&\mathbb I_2
 \end{pmatrix},\qquad \begin{pmatrix}
 \bm s_{I_j}\\s_{l_j}\\s_{r_j}
 \end{pmatrix}
 =
 \mathsf M_j
 \begin{pmatrix}
 \bm u_{I_j}\\v_{l_j}\\v_{r_j}
 \end{pmatrix}.
\end{equation}

Since each $\mathsf M_j$ acts only within a single
unit cell, the global transformation
$\mathsf M=\bigoplus_{j=1}^r\mathsf M_j$
is implemented by the depth-one local unitary
\begin{equation}
 U_{\mathsf M}\ket{\bm u,\bm v}
 =\ket{\mathsf M(\bm u,\bm v)}.
\end{equation}

Under this transformation, the intra-cell BPM matrix
becomes block diagonal:
\begin{equation}
 \begin{aligned}
 \mathsf M_j^{\mathsf T}
 A_{\rm cell}\mathsf M_j
 =
 \begin{pmatrix}
 \mathsf B_p&0\\
 0&A_{\rm end}
 +E_p^{\mathsf T}\mathsf B_p^{-1}E_p
 \end{pmatrix}.
 \end{aligned}
\end{equation}
Using Eq.~\eqref{cc:endpointidentity} and
$2\mu=0\pmod N$, the effective endpoint block reduces to
\begin{equation}
 A_{\rm end}
 +E_p^{\mathsf T}\mathsf B_p^{-1}E_p
 =
 \begin{pmatrix}
 0&-\kappa\\
 \kappa&0
 \end{pmatrix}.
\end{equation}
Thus, the diagonal endpoint terms cancel, while
an effective link of strength $\kappa$ is generated
between the two endpoints of each cell.
The original couplings between adjacent cells are
unchanged by the transformation and have the same
structure. Consequently, the internal sites decouple
from the endpoints, which form a uniform residual
chain with a zero-diagonal BPM matrix.

The full BPM matrix therefore factorizes as
\begin{equation}
 \mathsf M^{\mathsf T}A\mathsf M
 =
 \left(\bigoplus_{j=1}^r\mathsf B_p\right)
 \oplus A_{\mu=0},
\end{equation}
where $A_{\mu=0}$ is the $2r\times2r$
zero-diagonal BPM matrix acting on the endpoint
degrees of freedom. Correspondingly, the BPM
duality decomposes into
\begin{equation}
 U_{\mathsf M}^{\dagger}D_AU_{\mathsf M}
 =
 \left(\bigotimes_{j=1}^rD_{\mathsf B_p}\right)
 \otimes D_{A_{\mu=0}}.
 \label{cc:factorization}
\end{equation}

We next construct a BPM-invariant Hamiltonian
in the decoupled basis. Since $\mathsf B_p$ is
invertible, the duality $D_{\mathsf B_p}$
exchanges the two product states
\begin{equation}
 \begin{aligned}
 D_{\mathsf B_p}\ket{0}^{\otimes p}
 =\ket{+}^{\otimes p},D_{\mathsf B_p}\ket{+}^{\otimes p}
 =\ket{0}^{\otimes p}.
 \end{aligned}
 \label{cc:blockswap}
\end{equation}
It follows that the normalized symmetric combination
\begin{equation}
 \ket{\chi_p}
 =
 \frac{\ket{0}^{\otimes p}+\ket{+}^{\otimes p}}
 {\sqrt{2+2N^{-p/2}}}
\end{equation}
is invariant under $D_{\mathsf B_p}$.
We therefore introduce the local parent projector
\begin{equation}
 h_p=\mathbb I-\ket{\chi_p}\bra{\chi_p},
 \qquad
 D_{\mathsf B_p}\ket{\chi_p}=\ket{\chi_p}.
 \label{cc:blockstate}
\end{equation}

In the decoupled basis, consider the Hamiltonian
\begin{equation}
 H'=\sum_{j=1}^r(h_p)_j+(H_0)_{\rm end},
 \label{cc:finalH}
\end{equation}
where $(h_p)_j$ acts on the $p$ internal sites
of the $j$-th cell, and $(H_0)_{\rm end}$
is the BPM-invariant cluster Hamiltonian constructed
in the $\mu=0$ case, now acting on the residual endpoint chain.

All local terms in $H'$ mutually commute and
preserve the factorized BPM duality
$U_{\mathsf M}^{\dagger}D_AU_{\mathsf M}$.
Moreover, the Hamiltonian has a unique gapped
ground state
\begin{equation}
 \ket{\Omega'}
 =
 \ket{\chi_p}^{\otimes r}
 \otimes\ket{\mathrm{cluster}}_{\rm end}.
\end{equation}

The kernel symmetries are also preserved.
Indeed, the invertibility of each internal block
implies
\begin{equation}
 \ker(\mathsf M^{\mathsf T}A\mathsf M)
 =
 \{0\}^{pr}\oplus\ker A_{\mu=0}.
 \label{cc:kernelmapping}
\end{equation}
Hence, all kernel symmetries act exclusively
on the residual endpoint chain, where they
commute with $(H_0)_{\rm end}$, and therefore
with the full Hamiltonian $H'$.

Then mapping back to the original
qudit variables gives
\begin{equation}
 H=U_{\mathsf M}H'U_{\mathsf M}^{\dagger},
\end{equation}
with the unique gapped ground state
\begin{equation}
 \ket{\Omega}=U_{\mathsf M}\ket{\Omega'}.
\end{equation}
Since $U_{\mathsf M}$ is a depth-one local
unitary, $H$ remains local and has the same
spectrum as $H'$. Furthermore, by construction,
it preserves both the original BPM duality
and its kernel symmetries.

Finally, under the change of basis implemented by \(U_{\mathsf M}\), the kernel symmetry acts only on the endpoint degrees of freedom:
\[
U_{\mathsf M}^{\dagger}U_gU_{\mathsf M}
=\prod_j X_{l_j}^{g_1}X_{r_j}^{g_2}.
\]
The endpoint chain thus carries the same kernel symmetry action as in the \(\mu=0\) construction. Applying the same boundary fractionalization argument, we conclude that this phase also has kernel SPT index \(k=-\kappa\pmod N\).
\\~\\
\section{\large E: Details of effective qubit chain}
\label{app:toolbox}
This appendix proves the correspondence between operators of qudit chain and effective qubit chain. Throughout this appendix $(N,\kappa,\mu)=(4,1,2)$, $4\mid L$. The qudit clocks obey $ZX=\ii XZ$.

We have the definition
\begin{align}
 C_j&=Z_{j-1}X_jZ_{j+1},& K_j&=C_j^2=Z_{j-1}^2X_j^2Z_{j+1}^2=X_j^2B_j^2,\\
 M_j&=\frac{C_j+C_j^\dagger}{2},& R_j&=Z_j^2.
 \label{enc:operators}
\end{align}

Now, we can consider the operator as a mapping from the effective qubit chain to the subspace $K_j=1$ of the original  qudit chain :
\begin{equation}
 w_j\ket{s_{b,j}}=\frac{\ket{s_{d,j}=s_{b,j}}+\ket{s_{d,j}=s_{b,j}+2}}{\sqrt2},\qquad
 \cJ=\Ucl \otimes^L_{j=1} w_j.
 \label{enc:isometry}
\end{equation}
Here  $s_{b,j}=0,1$ corresponds to up and down configuration of a qubit and $s_{d,j}=0,1,2,3$ corresponds to the configuration of a qudit on the same $j$ site.  The cluster circuit
\begin{equation}
 \Ucl=\prod_j\mathrm{CZ}^{(4)}_{j,j+1},\qquad
 \mathrm{CZ}^{(4)}\ket{s,t}=\ii^{st}\ket{s,t}
\end{equation}
satisfies $\Ucl^{\dagger} C_j\Ucl=X_j$. Hence this reverse transformation $\cJ^\dagger$ first maps $K_j\to X^2_j$ by the cluster circuit and then builds a relation between qubit and qudit configuration under the constraint $K_j=1$.   

We can also check 
one has $\cJ^\dagger\cJ=1$, $\cJ\cJ^\dagger=\prod_j(1+K_j)/2$. The exact Pauli dictionary is
\begin{equation}
 \cJ^\dagger M_j\cJ=\cJ^\dagger C_j\cJ=\sigma_j^x,
 \qquad \cJ^\dagger R_j\cJ=\sigma_j^z.
 \label{enc:dictionary}
\end{equation}\\

\prlsection{Lower bound of $\Lambda$ for effective qubit chain} We now derive a sufficient lower bound on $\Lambda$ ensuring that
the ground state lies in the all-$K_j=+1$ sector, including at a
critical point.

Let $S=\{j:K_j=-1\}$ denote the set of encoding defects, and let
$H_S$ be the Hamiltonian restricted to the corresponding sector.
We compare its ground-state energy with that of the defect-free
Hamiltonian $H_{\varnothing}$ with all $K_j=1$. After conjugation by $\Ucl^\dagger$,
we have $K_j\mapsto X_j^2$, so each sector is a tensor product of
two-dimensional eigenspaces of $X_j^2$. These sectors can therefore
be identified with a common qubit Hilbert space, in which
\[
 R_j\mapsto\sigma_j^z,\qquad
 M_j\mapsto
 \begin{cases}
 \sigma_j^x,&j\notin S,\\
 0,&j\in S.
 \end{cases}
\]
In this representation, $H_S$ differs from $H_{\varnothing}$ only
through the deletion of certain interaction terms and the addition
of the defect penalty $\Lambda|S|$.

In particular, a defect at site $j$ removes the two adjacent
interchain couplings, $\eta M_{j-1}M_j$ and $\eta M_jM_{j+1}$.
Since each has operator norm $|\eta|$, deleting these terms can
lower the ground-state energy by at most $2|\eta|$. The total
possible energy reduction from the interchain couplings is
therefore bounded by $2|\eta||S|$. Counting the remaining
interaction terms in the same way gives the following upper
bounds on the total interaction norm that a single defect can
remove:
\begin{center}
\begin{tabular}{c|l}
Defect position & Upper bound on deleted interaction norm \\ \hline
$j$ odd
& $J_d+|J_x|+|J_z|+4|K|+2|\eta|$ \\
$j=0\pmod4$
& $2J_d+2|J_x|+3|K|+2|\eta|$ \\
$j=2\pmod4$
& $2|J_x|+3|K|+2|\eta|$
\end{tabular}
\end{center}
A term affected by several defects may be counted more than once,
which only weakens the bound. Thus the total possible energy
reduction from deleted interactions is at most $C_*|S|$, whereas
the penalty term contributes exactly $\Lambda|S|$. The variational
principle then gives
\begin{align}
 E_0(H_S)
 &\ge E_0(H_{\varnothing})+(\Lambda-C_*)|S|,
 \label{cone:defectcomparison}\\
 C_*
 &=\max\!\left\{
 J_d+|J_x|+|J_z|+4|K|,
 \;2J_d+2|J_x|+3|K|
 \right\}+2|\eta|.
 \label{cone:defectconstant}
\end{align}\\

\prlsection{Projection of the noninvertible symmetry}
 Now we rewrite the mapping \eqref{enc:isometry} in the qubit and qudit basis:
\begin{equation}
 \cJ\ket{\{s_{b,j}\}}=2^{-L/2}\sum_{\{u_j=0,1\}}
 \ii^{\Phi(\{s_{b,j}+2u_j\})}\ket{\{s_{d,j}=s_{b,j}+2u_j\}}
\end{equation}
where  $\Phi(\{s_j\})=\sum_js_js_{j+1}$.

Inserting it on both sides of the qudit BPM and summing over the two binary lift variables gives
\begin{equation}
 \bra{\{s'_{b,j}\}}\Deff\ket{\{s_{b,j}\}}=
 \ii^{\Phi(\{s_{b,j}\})-\Phi(\{s'_{b,j}\})-\sum_{j,k}s'_{b,j}A_{jk}s_{b,k}}
 \prod_j\delta_{s_{b,j-1}+s_{b,j+1}+s'_{b,j-1}+s'_{b,j+1},\,0\ ({\rm mod}\ 2)}.
\end{equation}
The delta functions ensure the binary difference $d_j=s_{b,j}+s'_{b,j}$ is constant on each sublattice of odd and even sites. There are four outputs: no flip, odd flip, even flip, and both flips. Their amplitudes are respectively $(-1)^{\sum s_{b,j}},1,1,(-1)^{\sum s_{b,j}}$; the last uses $4\mid L$. Hence, with
\begin{equation}
 P_o=\prod_{j\ {\rm odd}}\sigma_j^x,\qquad
 P_e=\prod_{j\ {\rm even}}\sigma_j^x,\qquad
 Q=\prod_j\sigma_j^z,
\end{equation}
the exact result is
\begin{equation}
 D\cJ=\cJ\Deff,\qquad
 \Deff=P_o+P_e+Q+P_oP_eQ.
 \label{enc:Deff}
\end{equation}
The cluster phase also gives
\begin{equation}
  U_a^{\rm eff}=\cJ^\dagger  U_a\cJ=P_oQ=\prod_j \sigma^y_{2j-1}\sigma^z_{2j},\qquad
  U_b^{\rm eff}=\cJ^\dagger  U_b\cJ=P_eQ=\prod_j \sigma^z_{2j-1}\sigma^y_{2j},
 \qquad \Deff=Q(1+  U_a^{\rm eff})(1+  U_b^{\rm eff}).
 \label{enc:ab}
\end{equation}

\label{enc:locality}
For a local Hamiltonian of the qubit chain, we can even have additional constraint on the effective symmetry:

Let $H=\sum_j h_j$ be a local qubit Hamiltonian, where each local $h_j$ is supported in the region $[j-r,j+r]$, where the finite $r$ is independent of $j$ and $L$. If $H$ commutes with $U_a^{\rm eff},U_b^{\rm eff},\Deff$, then for $L>4r+2$ it also commutes separately with $Q$. 

\textit{Proof.}
For brevity, set
\[
 F=(1+U_a^{\rm eff})(1+U_b^{\rm eff}),
 \qquad \Deff=QF.
\]
The internal symmetries imply $[H,F]=0$. Define the part of the Hamiltonian odd under conjugation by $Q$ as
\[
 H_-=\frac{H-QHQ}{2}
     =\sum_j h_{-,j},
 \qquad
 h_{-,j}=\frac{h_j-Qh_jQ}{2}.
\]
Since $[H,Q]=-2QH_-$, BPM invariance gives
\[
 \begin{aligned}
 0=[H,\Deff]=[H,QF]=[H,Q]F+Q[H,F]=-2QH_-F.
 \end{aligned}
\]
Multiplying on the left by $Q$ and using $Q^2=1$, we obtain
\begin{equation}
 \begin{aligned}
 0&=H_-(1+U_a^{\rm eff})(1+U_b^{\rm eff})\\
  &=H_-+H_-U_a^{\rm eff}
        +H_-U_b^{\rm eff}
        +H_-U_a^{\rm eff}U_b^{\rm eff}.
 \end{aligned}
 \label{enc:localityconstraint}
\end{equation}

Suppose each $h_j$ is supported on $X_j=[j-r,j+r]$, with $r$ independent of $L$. Because $Q$ acts onsite, $h_{-,j}$ is supported on the same region and admits the exact Pauli expansion
\[
 h_{-,j}
 =\sum_{S_{X_j}\in\mathcal \{1,\sigma^x,\sigma^y,\sigma^z\}^{\otimes |X_j|}}
 c_{j,S_{X_j}}S_{X_j},
 \qquad
 c_{j,S_{X_j}}
 =2^{-|X_j|}
 \operatorname{Tr}_{X_j}(S_{X_j}h_{-,j}),
\]
where the identity is understood outside $X_j$. Let $\operatorname{wt}(S)$ denote the number of nonidentity factors in a Pauli string $S$. Collecting identical strings across all local terms yields
\[
 H_-=\sum_{\operatorname{wt}(S)\le w}c_S S,
 \qquad w=2r+1.
\]

On the other hand, each of $ U_a^{\rm eff}$, $ U_b^{\rm eff}$ and $U_a^{\rm eff}U_b^{\rm eff}$
has a nonidentity Pauli factor at every site. For any Pauli string $S$, multiplying each global symmetry operator  leaves a nonidentity factor at every site outside $\operatorname{supp}(S)$. Consequently,
\[
 \operatorname{wt}(SG)
 \ge L-\operatorname{wt}(S)
 \ge L-w, \qquad G= U_a^{\rm eff},  U_b^{\rm eff}, U_a^{\rm eff} U_b^{\rm eff} .
\]
Thus the first term in \eqref{enc:localityconstraint} contains only strings of weight at most $w$, whereas the remaining three terms contain only strings of weight at least $L-w$.

For $L>2w=4r+2$, these two sets are disjoint. Using the orthogonality of the full-chain Pauli basis,
\[
 2^{-L}\operatorname{Tr}(SS')=\delta_{S,S'},
\]
we can extract the coefficient of any string $S$ with $\operatorname{wt}(S)\le w$ from \eqref{enc:localityconstraint}:
\[
 \begin{aligned}
 0
 &=2^{-L}\operatorname{Tr}
 \left[S H_-(1+U_a^{\rm eff})(1+U_b^{\rm eff})\right]\\
 &=2^{-L}\operatorname{Tr}(S H_-)
 =c_S.
 \end{aligned}
\]
All coefficients in $H_-$ therefore vanish, proving $H_-=0$ and hence $[H,Q]=0$.\\

\prlsection{Microscopic dictionary.}   Their leading
bosonization dictionary in the qubit chain is given by \cite{PhysRevB.99.075103,PhysRevB.99.165143}
\[
 m_z\sim\sin\theta,\qquad m_{\text{vbs}}\sim\cos\theta,
 \qquad m_x=\frac{1}{L}\sum_j\sigma^x_{2j}\sim\cos\phi,\qquad m_y=\frac{1}{L}\sum_j (-1)^j\sigma_{2j}^y\sim\sin\phi.
\]

The effective internal symmetries and inversion act as
\[
 \begin{array}{c|ccc}
 \text{continuum operator}&U^{\text{eff}}_b&U^{\text{eff}}_a/Q&I_0\\ \hline
 m_z(x)&-m_z(x)&m_z(x)&m_z(-x)\\
 m_{\text{vbs}}(x)&m_{\text{vbs}}(x)&m_{\text{vbs}}(x)&-m_{\text{vbs}}(-x)\\
 m_x(x)&-m_x(x)&-m_x(x)&m_x(-x)\\
 m_y(x)&m_y(x)&-m_y(x)&m_y(-x)
 \end{array}.
\]
This table then gives the symmetry transformation in the main text.\\

\section{\large  F: The stability of DQCP under derivative perturbations }
\label{cone:derivatives}
In this appendix, we discuss the stability under the spatial derivative perturbations as descendants of primary fields, which contributes oscillator excitation of boson CFT.

The CFT and primary fields are
\begin{align}
 H_0&=\frac{v}{2\pi}\int dx\,
 \left[\frac{g}{4}(\partial_x\phi)^2+g^{-1}(\partial_x\theta)^2\right],
 \label{cone:gaussian}\\
 V_{m,n}&\sim e^{\ii m\theta+\ii n\phi},\qquad
 \Delta_{m,n}=\frac{gm^2}{4}+\frac{n^2}{g},
 \qquad m,n\in\mathbb Z.
 \label{cone:dimensions}
\end{align}
The descendants of vertex
charge $(m,n)$ and total oscillator level $\ell$ have scaling dimension
\[
 \Delta_{m,n;\ell}
 =\frac{gm^2}{4}+\frac{n^2}{g}+\ell.
\]
The level $\ell$ counts the total number of derivatives, including derivatives
distributed among several factors.  For example,
$\partial_x^2V_{m,n}$ and
$:(\partial_x\theta)^2V_{m,n}:$ both have level two, whenever the
corresponding operator is nonzero.   Since
\[
\ell\ge2,\quad(m,n)\ne(0,0)
 \quad\Longrightarrow\quad\Delta_{m,n;\ell}>2.
\]
Every such operator with $\ell>1$ is irrelevant, even before applying the symmetry
selection rules.  It is precisely level one that can still have
$\Delta<2$ and requires the detailed classification below.

Since derivatives do not change the $U_a$ charge,
we can exclude every descendant of a
magnetically charged primary $e^{in\phi}$ with even $n\ne 0$.  The $U_b$-even combinations require the term to be a $\sin (n\theta)$ multiplying either current.
Since $\partial_x\theta$ is inversion-even and
$\partial_x\phi$ inversion-odd, the complete invariant
one-derivative families are
\begin{align}
 A_p&=(\partial_x\theta)\sin[(2p+1)\theta],
 &&p=0,1,\ldots,
 \label{cone:Aderivative}\\
 B_p&=\frac12\left\{\partial_x\phi,\sin(2p\theta)\right\},
 &&p=1,2,\ldots.
 \label{cone:Bderivative}
\end{align}

  All $A_p$ are spatial derivatives,
\begin{equation}
 A_p=-\frac{1}{2p+1}\partial_x\cos[(2p+1)\theta].
 \label{cone:spatialtotalderivative}
\end{equation}
Their integrals vanish with periodic boundary conditions and constant
bulk coefficients. 
The $B_p$ have dimensions $1+p^2g$.  Thus only
\begin{equation}
 B_1=\frac12\left\{\partial_x\phi,\sin(2\theta)\right\},
 \qquad \Delta_{B_1}=1+g,
 \label{cone:onlyderivative}
\end{equation}
can be relevant, namely when $g<1$.  At $g=1$ it is marginal by power
counting.  It is, however, a redundant perturbation at all these
values of $g$.

To demonstrate this beyond a linearized equation-of-motion argument,
first consider $H_\gamma=H_0+\gamma\int dx\,B_1$ and recall that
$\Pi=\partial_x\phi/\pi$.  Set $f(\theta)=\sin(2\theta)$.  Then
\[
 \gamma B_1=\frac{\pi\gamma}{2}\{\Pi,f\},\qquad
 \mathcal H_0=\frac{\pi vg}{2}\Pi^2
              +\frac{v}{2\pi g}(\partial_x\theta)^2.
\]
  Subtracting the extra square,
and then using $\sin^2(2\theta)=(1-\cos4\theta)/2$, gives
\begin{align}
 \frac{\pi vg}{2}\Pi^2+\frac{\pi\gamma}{2}\{\Pi,f\}
 &=\frac{\pi vg}{2}
       \left(\Pi+\frac{\gamma}{vg}f\right)^2
       -\frac{\pi\gamma^2}{2vg}f^2 \nonumber\\
 &=\frac{\pi vg}{2}
       \left(\Pi+\frac{\gamma}{vg}\sin2\theta\right)^2
       -\frac{\pi\gamma^2}{4vg}
       +\frac{\pi\gamma^2}{4vg}\cos4\theta.
 \label{cone:squarecompletion}
\end{align}
Thus the derivative interaction can be absorbed into a shifted
momentum, provided its order-$\gamma^2$ potential is retained.
The constant shifts the reference energy.
The induced $\cos4\theta$ term has dimension
\[
 \Delta[\cos4\theta]=4g>2\qquad\text{for }g>\tfrac12,
\]
which is therefore irrelevant when $1/2<g<2$.  Although
$B_1$ has naive dimension $1+g<2$ when $g<1$, it supplies no
additional independent relevant coupling.

\section{\large G: Entanglement spectrum and Boundary conformal field theory}
We now discuss the entanglement spectrum and its relation to the
energy spectrum of a boundary conformal field theory (BCFT).

\prlsection{Reduced density matrix under the qubit-to-qudit mapping.}
We consider the bipartition at $j_0\,|\,(j_0+1)=(4n+2)\,|\,(4n+3)$,
with the left and right half-chains denoted by $A$ and $B$,
respectively. Throughout this derivation, $\simeq$ denotes
equivalence under the transformations which preserve the nonzero spectrum of the reduced density matrix.

A normalized qubit state can be written as
\begin{equation}
|\psi_q\rangle
=
\sum_{\alpha,\beta,a,b}
(M_{ab})_{\alpha\beta}
|\alpha,a\rangle_A|b,\beta\rangle_B,
\qquad
\sum_{a,b}\operatorname{Tr}(M_{ab}M_{ab}^{\dagger})=1,
\end{equation}
where $a,b\in\{0,1\}$ label the physical
$\sigma^z=\pm 1$ states immediately to the left and right of the
cut, at sites $j=4n+2$ and $j=4n+3$, respectively. The indices
$\alpha,\beta$ label orthonormal bases for the remaining left and
right environments.

The qubit-to-qudit mapping in Eq.~\eqref{enc:isometry} is
implemented by the operator
$\prod_j\mathrm{CZ}^{(4)}_{j,j+1} \prod_j w_j$.
The corresponding qudit state is
\[
|\Psi_d\rangle=\prod_j\mathrm{CZ}^{(4)}_{j,j+1} \prod_j w_j|\psi_q\rangle.
\]
Undoing the gates supported entirely within either half-chain,
$U_A=\prod_{j,j+1\in A} \mathrm{CZ}^{(4)}_{j,j+1}$ and
$U_B=\prod_{j,j+1\in B}\mathrm{CZ}^{(4)}_{j,j+1}$, gives
\[
|\Psi'_d\rangle=U^{-1}_A U^{-1}_B|\Psi_d\rangle=\mathrm{CZ}^{(4)}_{4n+2,4n+3} \prod_j w_j|\psi_q\rangle.
\]
Since these transformations act separately within $A$ and $B$,
the resulting state has the same Schmidt weights as
$|\Psi_d\rangle$ across the chosen bipartition.

For the two boundary qudits, we identify each qudit with two qubits by writing
$s_{4n+2}=a+2u$ and $s_{4n+3}=b+2v$, where $u,v=0,1$ label
the $\sigma^z$ value of the auxiliary qubits.
The on-site map $w$ initializes these auxiliary qubits in
$\ket+$. For example,
\[
 w\ket{a}=\frac{\ket{s_{4n+2}=a}+\ket{s_{4n+2}=a+2}}{\sqrt2}=|a\rangle\otimes\frac{|u=0\rangle+|u=1\rangle}{\sqrt{2}} .
\]
Using
$
i^{(a+2u)(b+2v)}
=
i^{ab}(-1)^{ub+av}
$,
the crossing gate acts as
\begin{equation}
\mathrm{CZ}^{(4)}_{4n+2,4n+3}|a,+\rangle_A|b,+\rangle_B=
\frac{i^{ab}}{2}
\sum_{u,v}(-1)^{ub+av}
|a,u\rangle_A|b,v\rangle_B.
\end{equation}
We then apply a Hadamard transformation to each auxiliary factor:
\[
H_{u} \frac{|u=0\rangle+(-1)^b|u=1\rangle}{\sqrt{2}}=|b\rangle,~H_v\frac{|v=0\rangle+(-1)^a|v=1\rangle}{\sqrt{2}}=|a\rangle.
\]
The transformed boundary state is
\begin{equation}
H_u H_v\mathrm{CZ}^{(4)}_{4n+2,4n+3}|a,+\rangle_A|b,+\rangle_B=
i^{ab}|a,b\rangle_A|b,a\rangle_B.
\end{equation}
Thus, both boundary labels are recorded in orthogonal states on
both sides of the cut. The onsite Hadamard transformations also preserve the
Schmidt weights.

Then the resulting  state takes the form
\begin{equation}\label{eq: final state}
|\widetilde{\Psi}_d\rangle
=
\sum_{\alpha,\beta,a,b}
i^{ab}(M_{ab})_{\alpha\beta}
|\alpha,a,b\rangle_A|b,a,\beta\rangle_B.
\end{equation}
Here, $|\alpha\rangle$ and $|\beta\rangle$ now denote the
orthonormal environment  qudit states, obtained by applying the on-site
maps $w$ throughout the corresponding environments.
For example,
$|\alpha\rangle=\frac{|s_d=\alpha\rangle+|s_d=\alpha+2\rangle}{\sqrt{2}}$.

Tracing out $B$ eliminates all cross terms between distinct
boundary labels, yielding the exact block decomposition
\begin{equation}
\rho_A^d
\simeq
\bigoplus_{a,b=0,1} R_{ab},
\qquad
R_{ab}=M_{ab}M_{ab}^{\dagger}.
\end{equation}
We next consider the symmetries generated by
\[
U_a=\prod_j X_{4j+1}X^2_{4j+2}X_{4j+3},
\qquad
U_b^2=\prod_j X^2_{2j}.
\]
In the effective qubit chain, $U_a$ takes the form
\[
U_a^{\mathrm{eff}}
=
\prod_j \sigma^y_{2j-1}\sigma^z_{2j}
=
\mathcal L\otimes \sigma^z_{4n+2}
\otimes \sigma^y_{4n+3}\otimes \mathcal R,
\]
where $\mathcal{L}$ and $\mathcal R$ are unitary operators acting on the left
and right environment spaces, respectively.

For a qubit state satisfying
$U_a^{\mathrm{eff}}|\psi_q\rangle=\gamma|\psi_q\rangle$,
with $\gamma=\pm1$, comparison of the wavefunction coefficients
gives
\[
M_{a,1-b}
=
\gamma i(-1)^{a+b}\mathcal L M_{ab}\mathcal R^T,
\]
where we have used
$\sigma^z|a\rangle=(-1)^a|a\rangle$ and
$\sigma^y|b\rangle=i(-1)^b|1-b\rangle$.
Consequently,
\[
R_{a,1-b}
=
\mathcal L M_{ab}\mathcal R^T\mathcal R^*M_{ab}^{\dagger}\mathcal L^\dagger
=
\mathcal L R_{ab}\mathcal L^\dagger.
\]

Thus, for each fixed $a$, the two blocks labeled by $b=0,1$
are unitarily equivalent. After a further unitary transformation
within each block, the reduced density matrix takes the form
\begin{equation}
\rho_A^d
\simeq
R_{00}\oplus R_{00}\oplus R_{10}\oplus R_{10}.
\label{eq:RDM}
\end{equation}
Every nonzero eigenvalue of the reduced density matrix is
therefore at least twofold degenerate.

The role of $U_b^2$ is complementary. If the qudit state preserves
this on-site symmetry, tracing out $B$ gives
\[
[\rho_A^d,U_{b^2,A}]=0,
\qquad
U_{b^2,A}=\prod_{2j\in A}X_{2j}^2=Z_{4n+1}^2X_{4n+2}^2,
\]
where we use the constraint $Z^2_{2j-1}X^2_{2j}Z^2_{2j+1}=1$.

Under the above transformation, we have
\[
F_AU_{b^2,A} F^{-1}_{A}=\sigma^z_{u}
\]
where $F_A=H_u U_A^\dagger$. This gives $(-1)^b$ on the basis $|\alpha, a, u=b\rangle_A$ in Eq.~\eqref{eq: final state}.

Hence the block decomposition resolves the two eigenvalue
sectors of $U_{b^2,A}$, while $U_a$ exchanges these sectors. Their anti-commutation relation ensure the twofold degeneracy.

~\\

\prlsection{Corresponding boundary conformal field theory.} As discussed in Ref.~\cite{PhysRevLett.132.086503}, in the
finite-entanglement scaling regime, the low-lying entanglement
spectrum of an infinite MPS with finite bond dimension can be
related, up to an overall shift and scale, to the energy spectrum
of a BCFT with a physical boundary $B_p$ and an entanglement
boundary $B_e$.

The entanglement boundary is determined by the microscopic
factorization and regularization at the cut used to define the
reduced density matrix. In the block decomposition
of Eq.~\eqref{eq:RDM}, $a=0,1$ label the two eigenstates of
$\sigma^z_{4n+2}$. Motivated by the continuum correspondence
$\sigma^z_{4n+2}\sim \sin\theta$, the entanglement boundary conditions are
\begin{equation}
B_{e,0}\ \longrightarrow\ D_\theta(\pi/2),\qquad
B_{e,1}\ \longrightarrow\ D_\theta(-\pi/2).
\label{eq:RGassignment}
\end{equation}
Here, $D_\theta(\delta)$ denotes a Dirichlet boundary condition
with $\theta=\delta$.

The physical boundary, by contrast, is determined by the
effective relevant boundary perturbation associated with finite
bond dimension. We consider MPS that preserve the
unbroken symmetries of the FM or VBS phase. Then their physical boundaries are
\begin{equation}
B_p^{\rm FM}=D_\theta(\pi/2)\ \text{or}\ D_\theta(-\pi/2),
\qquad
B_p^{\rm VBS}=D_\theta(0)\ \text{or}\ D_\theta(\pi).
\label{eq:Bp}
\end{equation}
\\

\section{\large H: Additional numerical results }
\begin{figure}[t]
\centering
\includegraphics[width=1.05\textwidth]{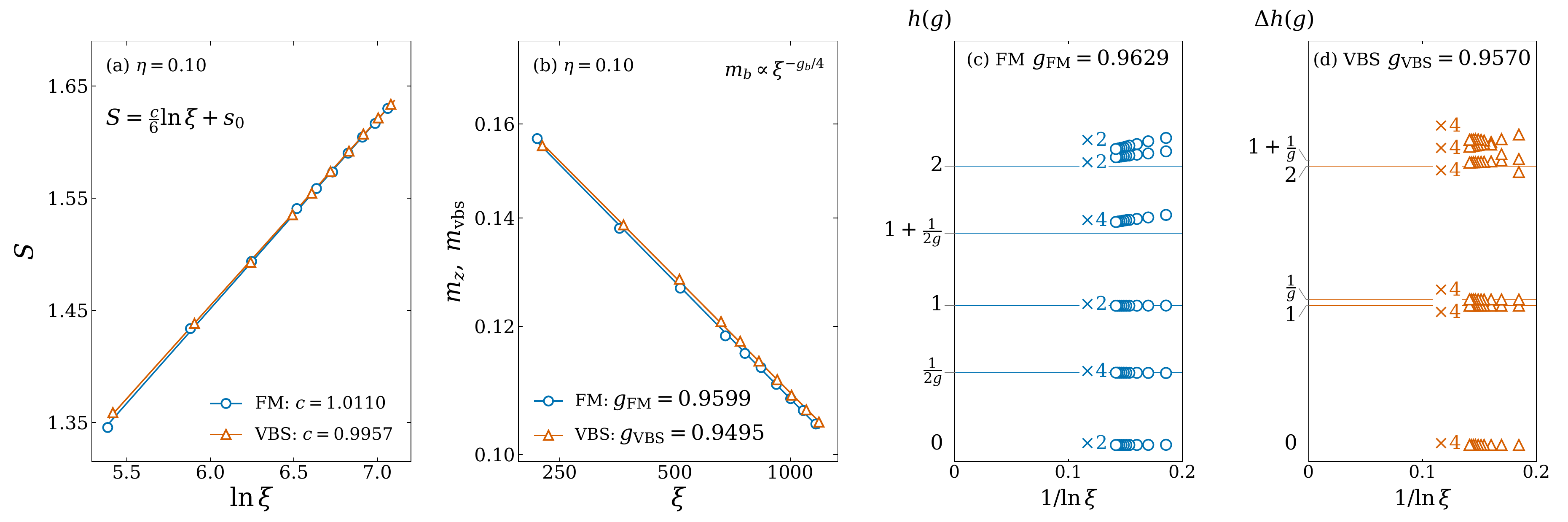}
\caption{Numerical results for DQCP at $\eta=0.1$ (a) Entanglement entropy $S$ as a function of the correlation length $\xi$. (b) Log–log plots of the FM and VBS order parameters, respectively, as functions of $\xi$. Power-law fits at $K_c$ yield the scaling dimensions $\Delta$, from which we extract the Luttinger parameter $g$ using $\Delta=g/4$. (c-d) Entanglement spectra at the DQCP for two iMPS ansätze that preserve the unbroken symmetries of the FM and VBS phases, respectively. Each spectrum is shifted so that the lowest level lies at $h=0$ and rescaled using the reference excited level at $h=1$. The Luttinger parameter $g$ is extracted from the excited levels at $h=1/(2g)$ and $h=1/g$.}
\label{fig:DQCP eta010}
\end{figure}

\begin{figure}[t]
\centering
\includegraphics[width=1.05\textwidth]{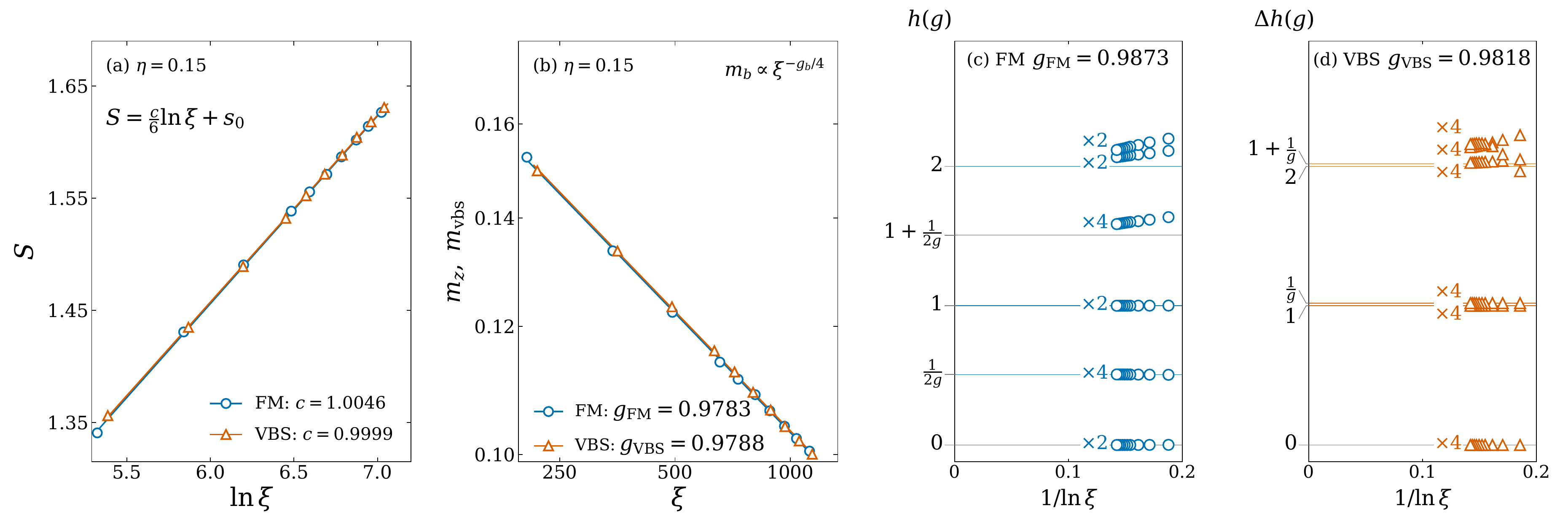}
\caption{Numerical results for DQCP at $\eta=0.15$ (a) Entanglement entropy $S$ as a function of the correlation length $\xi$. (b) Log–log plots of the FM and VBS order parameters, respectively, as functions of $\xi$. Power-law fits at $K_c$ yield the scaling dimensions $\Delta$, from which we extract the Luttinger parameter $g$ using $\Delta=g/4$. (c-d) Entanglement spectra at the DQCP for two iMPS ansätze that preserve the unbroken symmetries of the FM and VBS phases, respectively. Each spectrum is shifted so that the lowest level lies at $h=0$ and rescaled using the reference excited level at $h=1$. The Luttinger parameter $g$ is extracted from the excited levels at $h=1/(2g)$ and $h=1/g$.}
\label{fig:DQCP eta015}
\end{figure}

Finally we show other results, including the order parameter for $0.53<K<0.57$ and $\eta=0.05$ and the information of DQCP with $\eta=0.1$ and $\eta=0.15$, as shown in Fig.~\ref{fig:order parameter}, Fig.~\ref{fig:DQCP eta010} and Fig.~\ref{fig:DQCP eta015}.
\begin{figure}[t]
\centering
\includegraphics[width=0.6\textwidth]{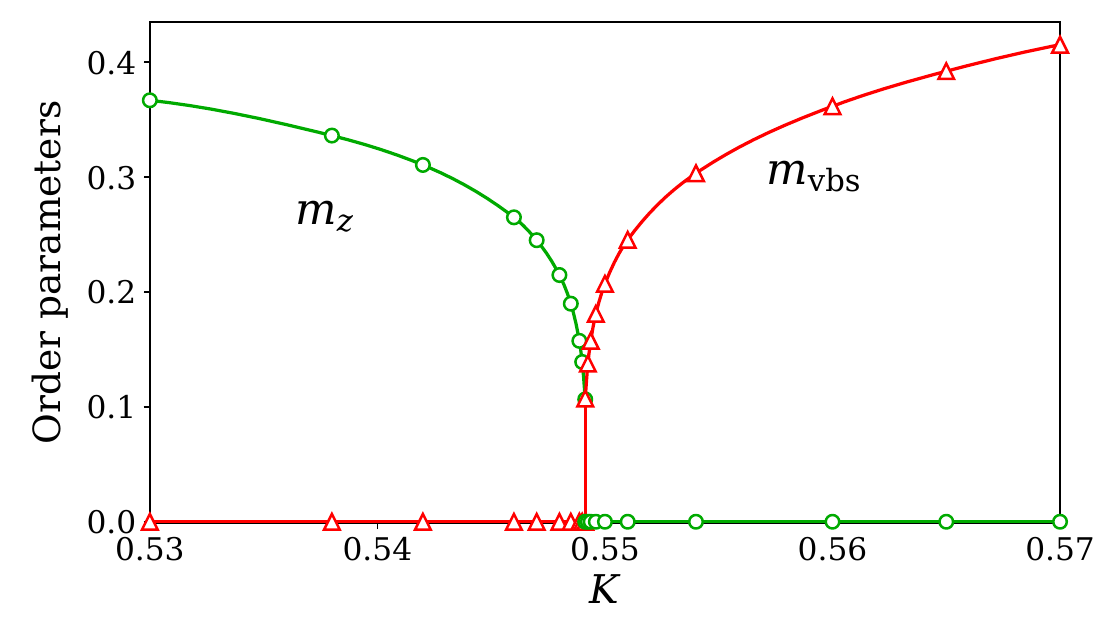}
\caption{Order parameters for $0.53<K<0.57$ and $\eta=0.05$}
\label{fig:order parameter}
\end{figure}


\end{document}